\documentclass[a4paper,11pt]{article}
\pdfoutput=1 % if your are submitting a pdflatex (i.e. if you have
\usepackage{jcappub} % for details on the use of the package, please
\usepackage[T1]{fontenc} % if needed

\newcommand{\lb}{\left(} 
\newcommand{\rb}{\right)}
\newcommand{\lsb}{\left[} 
\newcommand{\rsb}{\right]} 
\newcommand{\lcb}{\left\{} 
\newcommand{\rcb}{\right\}} 
\newcommand{\lan}{\left\langle} 
\newcommand{\ran}{\right\rangle} 

\title{\boldmath Computing First-Passage Times with the Functional Renormalisation Group}

 \author{G. Rigopoulos}
 \author{A. Wilkins}
 \affiliation{School of Maths, Stats \& Physics, \\Newcastle University,\\Newcastle, UK}

\emailAdd{gerasimos.rigopoulos@newcastle.ac.uk}
\emailAdd{a.wilkins3@newcastle.ac.uk}
\abstract{We use Functional Renormalisation Group (FRG) techniques to analyse the behaviour of a spectator field, $\sigma$, during inflation that obeys an overdamped Langevin equation. We briefly review how a derivative expansion of the FRG can be used to obtain Effective Equations of Motion (EEOM) for the one- and two-point function and derive the EEOM for the three-point function. We show how to compute quantities like the amplitude of the power spectrum and the spectral tilt from the FRG. We do this explicitly for a potential with multiple barriers and show that in general many different potentials will give identical predictions for the spectral tilt suggesting that observations are agnostic to localised features in the potential. Finally we use the EEOM to compute first-passage time (FPT) quantities for the spectator field. The EEOM for the one- and two-point function are enough to accurately predict the average time taken $\lan \mathcal{N}\ran$ to travel between two field values with a barrier in between and the variation in that time $\delta \mathcal{N}^2$. It can also accurately resolve the full PDF for time taken $\rho (\mathcal{N})$, predicting the correct exponential tail. This suggests that an extension of this analysis to the inflaton can correctly capture the exponential tail that is expected in models producing Primordial Black Holes.}

\begin{document}
\maketitle
\flushbottom

\section{\label{sec:RGSpec_intro} Introduction}
Stochastic Inflation \cite{Starobinsky1988, Nambu1988,Nambu1989,Mollerach1991,Salopek1991,Habib1992, Linde1994,Starobinsky1994} is the leading framework to describe the evolution of non-linear perturbations and their backreaction on the dynamics of the inflaton, $\phi$. The basic idea is to split inflationary perturbations into short- and long-wavelength components. The long-wavelength perturbations can be treated as effectively classical greatly simplifying the analysis. The initially short-wavelength quantum perturbations are stretched by the rapid inflationary expansion and can be consistently included as a classical random noise term on the dynamical equations which is a well established approximation for the the behaviour of IR quantum fields in inflationary spacetimes \cite{Tsamis2005,Finelli2009,Finelli2010,Garbrecht2014,Garbrecht2015,Moss2017} -- see \cite{Cable2021,Cable2022} however for how this picture breaks down for too massive test fields and see \cite{Cohen2021b} for next-to-next-to leading order corrections to the standard stochastic framework. In this way it is clear that the stochastic framework can be imagined as an Effective Field Theory (EFT) of the long-wavelength sector.\\

The stochastic nature of the dynamics means that the time taken (measured in e-folds) for the inflaton to reach the end of inflation, corresponding to $\phi_e$, is also a stochastic quantity, denoted by $\mathcal{N}$. The stochastic $\delta \mathcal{N}$ formalism \cite{Enqvist2008, Fujita2013, Fujita2014, Vennin2015} allows one to compute the coarse-grained comoving curvature perturbation on uniform energy density time-slices $\mathcal{R}_{cg}$ through:
\begin{eqnarray}
\mathcal{N} - \left\langle \mathcal{N}\right\rangle = \mathcal{R}_{cg} = \dfrac{1}{(2\pi)^{3/2}}\int_{k_{in}}^{k_{end}}\mathrm{d}\vec{k}\mathcal{R}_{\vec{k}}e^{i\vec{k}\cdot\vec{x}} \label{eq:Rcg_defn}
\end{eqnarray}  
which -- as the name suggests -- is just the usual comoving curvature perturbation coarse-grained between scales $k_{in}$, the scale that crossed the Hubble radius at initial time, and $k_{end}$, the scale that crosses out the Hubble radius at final time. This reduces the problem of computing curvature perturbations to the one of performing first-passage time (FPT) analysis on the stochastic equations of motion to obtain the PDF for exit time $\rho (\mathcal{N})$. This formalism is most useful when considering large perturbations that might collapse to form Primordial Black Holes -- see e.g. \cite{Pattison2017,Pattison2021,Rigopoulos2021,Animali2022}. It is however very difficult to accurately resolve the tail of $\rho (\mathcal{N})$ for all but the simplest potentials by direct numerical simulation as this is usually prohibitively expensive -- see \cite{Figueroa2020,Figueroa2021} for a treatment of this problem and \cite{Jackson2022} for a possible workaround using importance sampling.\footnote{While preparing this manuscript \cite{Tomberg2022} appeared on the arXiv which claims to offer an alternative method which is computationally cheap.} \\

In this work we present an alternative method for computing FPT quantities that utilises an effective description of the stochastic dynamics developed in \cite{Wilkins2021}. As a proof of concept we will focus on a spectator scalar field $\sigma$ that is present during inflation. This effective description captures the aggregate effect of fluctuations embodied in an \emph{effective action} $\Gamma[\Sigma(\alpha)]$, a functional of the average field value $\Sigma(\alpha) \equiv \langle \sigma(\alpha) \rangle$  where $\alpha$ is the number of e-folds. The effective action can be thought of as an analogue to the statistical free energy and can be derived from the partition function or generating functional via a Legendre transform. Once obtained, the effective action can be used to compute $n$-point correlation functions of the field values $\langle \sigma(\alpha_1) \sigma(\alpha_2)\ldots \sigma(\alpha_n) \rangle $, characterising the system's statistical properties. These n-point functions will allow us to compute the entire $\rho (\mathcal{N})$ distribution. To obtain this effective action we will be coarse-graining the system in time such that we obtain e.g. an effective potential that incorporates the quantum fluctuations on all timescales. To achieve this we will use a technique known as the exact, or non-perturbative, or Functional Renormalisation Group (FRG) \cite{Wetterich1993, Morris1994} as applied to Brownian motion \cite{Duclut2017,Wilkins2021} -- see \cite{Berges2002} for a review and an entry point to the literature on the FRG, \cite{Dupuis2020} for a comprehensive overview of applications as well as e.g. \cite{Gies2012, Delamotte2012} for more elementary introductions. \\

We begin in section \ref{sec:Stoch_Spec} by reviewing the concept of a stochastic spectator in the early universe and outline how this behaviour can be related to the path integral formulation for stochastic behaviour \cite{Martin1973,DEDOMINICIS1976,Janssen1976,DeDominicis1978} -- see also \cite{Wilkins2021} for a pedagogical overview. In section \ref{sec:RG_Stoch_Spec} we adapt the Effective Equations of Motion (EEOM) for the one- and two-point functions developed in \cite{Wilkins2021} to a spectator field and introduce the EEOM for the third central moment. In section \ref{sec:Cosmo_Spec} we discuss how we can obtain cosmological observables like the power spectrum and spectral tilt from FRG computed quantities. In section \ref{sec:FPT_spec} we show how the FRG can solve the FPT problem for a spectator field and can predict quantities such as $\lan \mathcal{N} \ran $ and $\delta \mathcal{N}^2$ for two complicated potentials. We leave extending the analysis of section \ref{sec:FPT_spec} beyond a normal distribution to the appendix.

\section{\label{sec:Stoch_Spec} The Stochastic Spectator}
The simplest models of inflation assume that it is driven by a single scalar field called the inflaton, $\phi$. However there is reason to suspect that other scalar fields would be present during the inflationary period. For instance string theory predicts the presence of many extra light moduli fields \cite{Turok1988, Damour1996,Kachru2003} and unless we are dealing with Higgs inflation -- see e.g. \cite{Martin2014} -- we would expect the Higgs field to be present also. \\

To be concrete we introduce another scalar field $\sigma$ evolving in a potential $U(\sigma)$. Then the first Friedmann equation reads:
\begin{eqnarray}
3M_{p}^2 H^2 = \dfrac{1}{2}\dot{\phi}^2 + V(\phi) + \dfrac{1}{2}\dot{\sigma}^2 + U(\sigma) \label{eq:Friedmann_sigma}
\end{eqnarray}
so that there are also contributions from the kinetic and potential energy of $\sigma$. We also obtain a Klein-Gordon equation for $\sigma$:
\begin{eqnarray}
\ddot{\sigma} +3H\dot{\sigma} + \dfrac{\mathrm{d}U(\sigma)}{\mathrm{d}\sigma} = 0 \label{eq:K-G_sigma}
\end{eqnarray}
If the energy scale of $\sigma$ is comparable to $\phi$ -- i.e. $V(\phi) \sim U(\sigma)$ -- then both fields are relevant for the dynamics of inflation and we are in a multi-field inflation scenario. Inflation will then proceed along a direction in the $(\phi, \sigma)$ field space -- see \cite{Pinol2021} for an example of how to deal with this in the stochastic inflation approach. In this paper we will consider the much more straightforward scenario where $U(\sigma) \ll V(\phi)$ such that the full Friedmann equation (\ref{eq:Friedmann_sigma}) is well described by the standard single-field Friedmann equation:
\begin{eqnarray}
3M_{p}^2 H^2 = \dfrac{1}{2}\dot{\phi}^2 + V(\phi) \label{eq:Friedmann_Inflation}
\end{eqnarray}
and the field $\sigma$ does not affect the inflationary dynamics. It is therefore referred to as a \emph{spectator} field. \\

Despite not directly affecting inflationary dynamics spectator fields can be very important. In the \emph{curvaton} scenario \cite{Linde1997,Moroi2001,Lyth2002,Moroi2002,Lyth2003,Lyth2005,Vennin2015a} the inflaton produces a subdominant contribution to the primordial density perturbation and the spectator field is the main contribution to the curvature perturbation hence the name. This is typically achieved by having the inflaton decay into radiation before the curvaton decays so that there is a period where the curvaton is the dominant contribution to the energy budget. In some cases this can even drive a short second period of inflation -- see \cite{Vennin2015a} for a full breakdown of all the possible configurations. A curvaton field could also be used as a means of measuring the duration of inflation \cite{Torrado2018}. Even if the spectator field is not the dominant contribution to the curvature perturbation observed in the CMB a spectator field could still form PBHs from field bubbles \cite{Maeso2022}. An inflationary period also affects the dynamics of any spectator field and if this field becomes important later on (e.g. the Higgs) it is useful to know how inflation sets the initial conditions for these spectator fields after inflation is over. 

\subsection{Coarse-graining a spectator field}
We can split the spectator into long, $\sigma_{\scalebox{0.5}{$>$}}$, and short, $\sigma_{\scalebox{0.5}{$<$}}$, wavelength modes and inflation will force the short wavelength modes to backreact on the long wavelength modes. The key difference as compared to the inflaton is that as this field is a pure spectator this backreaction does not modify the \emph{geometry} of the background spacetime. Assuming we are dealing with overdamped motion (i.e. the slow-roll limit) it is straightforwardly shown -- see e.g. the original treatments \cite{Starobinsky1988,Starobinsky1994} -- that the equation of motion for $\sigma_{\scalebox{0.5}{$>$}}$ is:
\begin{eqnarray}
\dfrac{\mathrm{d}\sigma_{\scalebox{0.5}{$>$}}}{\mathrm{d}\alpha} &=& -\dfrac{1}{3H^2}\dfrac{\partial U(\sigma_{\scalebox{0.5}{$>$}})}{\partial \sigma_{\scalebox{0.5}{$>$}}} + \eta (\alpha) \label{eq:Langevin_sigma}\\
\lan \eta (\alpha) \eta (\alpha ')\ran &=& \dfrac{H^2}{4\pi^2} \delta (\alpha - \alpha ')\label{eq:sigma_noise}
\end{eqnarray}
where we have set $M_{\mathrm{p}}^2 =1$. In principle the value of the Hubble parameter will vary with time depending on the inflationary potential. We will choose the background inflationary potential to be of the plateau type so that $H$ is roughly constant and we can therefore assume that the spectator field exists in an exact de Sitter background. While the noise term in (\ref{eq:sigma_noise}) has been computed in the massless limit, it has been shown recently \cite{Cable2021,Cable2022} how to modify the noise appropriately for more massive spectator fields. We will assume here that the field is sufficiently light such that (\ref{eq:sigma_noise}) is a good approximation, in any case the procedure we outline in this chapter is easily adapted to incorporate different values of the noise. To lighten the notation we will drop the subscript on $\sigma_{\scalebox{0.5}{$>$}}$ going forward and $\sigma$ can be assumed to refer to the coarse-grained long-wavelength field. We now introduce a reference Hubble scale $H_0$\footnote{Not to be confused with the value of the Hubble parameter today which is also often called $H_0$.} to define the dimensionless Hubble parameter $\hat{H}$ and in turn the other terms in equation (\ref{eq:Langevin_sigma}):
\begin{subequations}
\begin{align}
H &=  H_0 \hat{H} \label{eq:H_dimless}\\
	\sigma &=  \dfrac{H_0}{2\pi}\hat{\sigma} \label{eq:sigma_dimless} \\
	U(\sigma) &=  \dfrac{3\hat{H}^2H_0^4}{4\pi^2}\hat{U}(\hat{\sigma}) \label{eq:V_dimless}\\
	\eta (\alpha) &=  \dfrac{H_0}{2\pi}\hat{\eta} (\alpha )  \label{eq:eta_dimless}
\end{align}\label{eq:all_sigma_dimensionless}   
\end{subequations}
Notice that as the number of e-folds $\alpha$ is already dimensionless we do not need to rescale it. Also worth commenting on is that the dimensionless potential $\hat{U}$ depends on the dimensionless Hubble parameter $\hat{H}$. In the standard Brownian motion system the temperature of the system determines the amplitude of the (thermal) fluctuations but the amount of damping is determined by another parameter. In our case the friction coefficient and amplitude of the noise are both determined by the same parameter $H$ hence why our dimensionless potential effectively depends on the \textit{temperature} of the system. We will deal with this more when we come to section \ref{sec:RG_Stoch_Spec}. All this will give us the following dimensionless Langevin equation:
\begin{eqnarray}
\dfrac{\mathrm{d}\hat{\sigma}}{\mathrm{d}\alpha} &=& -\dfrac{\partial \hat{U}(\hat{\sigma})}{\partial \hat{\sigma}} + \hat{\eta} (\alpha) \label{eq:Langevin_sigma_dimless}\\
\lan \hat{\eta} (\alpha) \hat{\eta} (\alpha ')\ran &=&\hat{H}^2 \delta (\alpha - \alpha ')\label{eq:sigma_noise_dimless}
\end{eqnarray}
which we can readily identify with the thermal dimensionless Langevin equation from \cite{Wilkins2021} by making the transformation $\hat{H}^2 \rightarrow \Upsilon$. We will now drop the hat on $\sigma $ for notational simplicity. 
\subsection{The Spectator Path Integral}
This stochastic problem can be reformulated in terms of a path integral \cite{Martin1973,DEDOMINICIS1976,Janssen1976,DeDominicis1978}. Modifying the Brownian Path Integral \cite{Wilkins2021} to be in terms of our new dimensionless parameters yields:
\begin{subequations}
\begin{align}
\mathcal{P}(\sigma_{f}\vert \sigma_{i}) &= \int\mathcal{D}\sigma\mathcal{D}\tilde{\sigma} \mathcal{D}c\mathcal{D}\bar{c} \text{ exp}\left[-\mathcal{S}_{Spect}(\sigma,\tilde{\sigma},\bar{c},c)\right] \label{eq:TransProb_Sigma} \\
\mathcal{S}_{Spect}(\sigma,\tilde{\sigma},\bar{c},c) &= \int \text{d}\alpha\bigg[ \frac{\hat{H}^2}{2}\tilde{\sigma}^2 - i\tilde{\sigma}(\partial_{\alpha}{\sigma}+ \hat{U}_{,\sigma}) - \bar{c}\left( \partial_{\alpha} +  \hat{U}_{,\sigma\sigma} \right)c \bigg] \label{eq: PDF action_Sigma}    
\end{align}
\end{subequations} 
where we have introduced the response field $\tilde{\sigma}$ to $\sigma$ and the anticommuting variable $c$ and $\bar{c}$. We can then in analogy with \cite{Wilkins2021} make the following identifications:
\begin{subequations}
\begin{align}
\sigma(\alpha) &\equiv  \hat{H}\,\varphi(\alpha) \\
	\hat{U}(\sigma) &\equiv  {\hat{H}^2} \, W(\varphi)  \\
	\tilde{\sigma} &\equiv  \dfrac{1}{\hat{H}}\,(i\partial_{\alpha}{\varphi} - \tilde{F}) \\
	\bar{c}c &\equiv  i\bar{\rho}\rho   
\end{align}\label{eq:sigma_identifications}   
\end{subequations}
We can then say that the spectator action (\ref{eq: PDF action_Sigma}) is equivalent to the action for Euclidean Supersymmetric Quantum Mechanics:
\begin{eqnarray}
	\mathcal{S}_{Spect}[\varphi, \tilde{F}, \bar{\rho}, \rho] = \left[W(\varphi_f) -  W(\varphi_i)\right] + \mathcal{S}_{SUSY}
\end{eqnarray}
where
\begin{eqnarray}
	\mathcal{S}_{SUSY}[\varphi, \tilde{F}, \bar{\rho}, \rho] = \int dt\bigg[\dfrac{1}{2}\dot{\varphi}^2 + \dfrac{1}{2}\tilde{F}^2 + i\tilde{F}W_{,\varphi}(\varphi) -i \bar{\rho}(\partial_{t} +  W_{,\varphi\varphi}(\varphi))\rho\bigg]  \label{eq:SUSYClass}
\end{eqnarray} 
up to a factor depending on the initial and final field values $\sigma_i$ \& $\sigma_f$; these terms can be simply taken outside the path integral as an exponential prefactor. In the SUSY action (\ref{eq:SUSYClass}) $\rho$ \& $\bar{\rho}$ are the fermionic fields and $\varphi$ \& $\tilde{F}$ are the bosonic fields \cite{Synatschke2009}. As the dynamics of Brownian motion and supersymmetry (SUSY) are essentially equivalent one can exploit the FRG flow equations derived in \cite{Synatschke2009} and apply them to a stochastic system \cite{Wilkins2021}. We outline how to do this in the context of a spectator field in the next section.

\section{\label{sec:RG_Stoch_Spec} The Effective Equations of Motion for a Spectator Field}
A standard formulation of classical mechanics involves the principle of least action. If one considers the classical action $\mathcal{S}$:
\begin{eqnarray}
\mathcal{S} = \int dt~ L(x,\dot{x})
\end{eqnarray}
where $L(x,\dot{x})$ is the Lagrangian, then one can obtain the equations of motion by taking the variational derivative and setting it equal to zero:
\begin{eqnarray}
\dfrac{\delta \mathcal{S}}{\delta x} = 0 \label{eq:delta S = 0}
\end{eqnarray}
The Effective Action (EA) $\Gamma$ is so named because its definition makes it look like a classical action but includes the effect of fluctuations that have been integrated out. Defining 
\begin{equation}
	e^{\mathcal{W}[{\mathcal{J}}]} = \int\mathcal{D}\Vec{\sigma} ~ e^{-\mathcal{S}[\Vec{\sigma}]+\int d\alpha  {\mathcal{J}}\Vec{\sigma}}
\end{equation}
where $\Vec{\sigma} = [\sigma, \tilde{\sigma}, \bar{c},c]$ and $\mathcal{J}$ is a collection of external sources. The effective action $\Gamma[\Vec{\Sigma}]$ is then defined as the Legendre transform of $\mathcal{W}[{\mathcal{J}}]$:
\begin{equation}
	\Gamma[\Vec{\Sigma}] = \int_\alpha {\mathcal{J}} \Vec{\Sigma} - \mathcal{W}[{\mathcal{J}}]
\end{equation}
where
\begin{equation}
	\Vec{\Sigma} = \left\langle \Vec{\sigma}  \right\rangle = \dfrac{\int\mathcal{D}\Vec{\sigma} ~\Vec{\sigma} e^{-\mathcal{S}[\Vec{\sigma}]}}{\int\mathcal{D}\Vec{\sigma} ~ e^{-\mathcal{S}[\Vec{\sigma}]}}\,.
\end{equation}
We then have
\begin{eqnarray}
	\dfrac{\delta \Gamma}{\delta \Vec{\Sigma}}= {\mathcal{J}}
\end{eqnarray}
Therefore $\Gamma$, the central object of the FRG, leads to effective equations of motion that incorporate the aggregate effects of the effectively thermal, quantum fluctuations.

%The Effective Action (EA) $\Gamma$ is so named because its definition makes it look like a standard classical action once fluctuations have been integrated out:
%\begin{eqnarray}
%e^{-\Gamma [\Sigma,\tilde{\Sigma},\bar{C},C ]} = \int\mathcal{D}\sigma \mathcal{D}\tilde{\sigma} \mathcal{D}\bar{c}\mathcal{D}c~e^{-\mathcal{S}[\sigma, \tilde{\sigma}, \bar{c},c]}
%\end{eqnarray}
%In \cite{Wilkins2021} it was shown how to extend the variational principle used to obtain the classical equations of motion from $S$ to obtain \textit{effective equations of motion} (EEOM) from $\Gamma$. As the FRG has $\Gamma$ as its central object it is ideally placed to calculate these EEOM. Also of importance is the fact that $\Gamma$ is written in terms of the mean fields
%\begin{subequations}
%\begin{align}
%\Sigma &=  \lan \sigma\ran_{J_{\sigma}} \\
%\tilde{\Sigma} &=  \lan \tilde{\sigma}\ran_{J_{\tilde{\sigma}}} \\
%C &=  \lan c \ran_{\vartheta} \\
%\bar{C} &=  \lan \bar{c}\ran_{\bar{\vartheta}}    
%\end{align}   \label{eq:Spect_mean_fields} 
%\end{subequations}
%directly. These in turn depend on the currents:
%\begin{equation}
% \int \mathrm{d}\alpha \,{\mathcal{J}} \Vec{\Sigma} \equiv \int \mathrm{d}\alpha \left(  J_\sigma \sigma + J_{\tilde{\sigma}} \tilde{\sigma} + \bar{c}\vartheta + \bar{\vartheta}c    \right) \label{eq:currents_spect}
%\end{equation}
In \cite{Wilkins2021} it was shown how to extend the variational principle used to obtain the classical equations of motion from $S$ to obtain \textit{effective equations of motion} (EEOM) from $\Gamma$. With this in mind you can obtain the EEOM for the average field value $\Sigma$ by taking the appropriate variational derivative of the EA:
\begin{eqnarray}
\dfrac{\delta\Gamma}{\delta \Sigma(\alpha)} = 0 \label{eq:gen QEOM}
\end{eqnarray}
with no external sources. Before we can practically take variational derivatives of the EA we need to first compute it using the Functional Renormalisation Group.

\subsection{\label{sec:FRG}The Functional Renormalisation Group for Brownian Motion}
In this subsection we will briefly review a particular formulation of the Renormalisation Group known as the Functional Renormalisation Group (FRG) \cite{Wetterich1993, Morris1994} in the context of Brownian motion \cite{Duclut2017,Wilkins2021} . It has many advantages over the original Wilsonian treatment, the most obvious is its ability to handle systems with strong couplings. As the name suggests, the FRG uses \textit{functional methods} to describe the computation correlation functions of the fields. This is typically done through the use of \textit{generating functionals} which in principle should contain all relevant physical information about a theory. \\

\begin{figure}[t!]
    \centering
    \includegraphics[width = 0.95\linewidth]{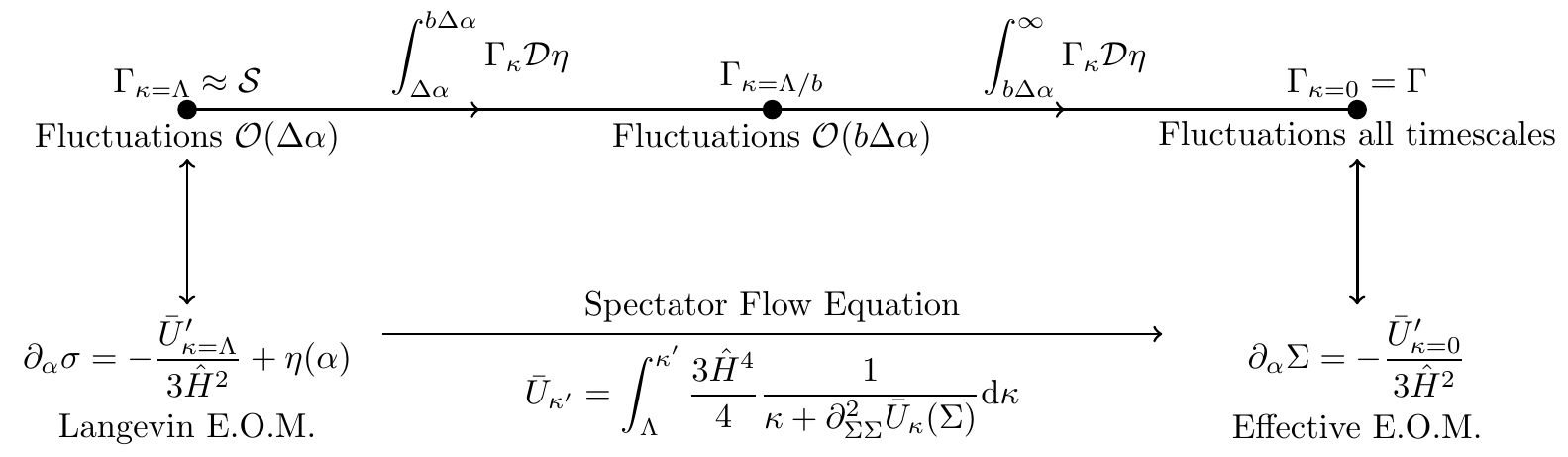}
    \caption{A schematic diagram of how the FRG takes the Langevin equation for a spectator field (\ref{eq:Langevin_sigma_dimless}) and creates an effective theory that incorporates the effect of fluctuations. The integration over $\eta$ in the top line should be understand as integrating out all fluctuations whose wavelength \textit{in time} -- i.e. how often they occur -- lies between the two integration bounds.}
    \label{fig:fRG_flow_schematic_EEOM}
\end{figure}

In line with \cite{Wilkins2021} we will focus on the derivative expansion which assumes that the EA can be written in the same functional form as the classical action (\ref{eq: PDF action_Sigma}) with the caveat that some terms will now depend on the renormalisation scale $\kappa$. In this way we have introduced a new object $\Gamma_{\kappa}$ known as the Regulated Effective Action (REA) that depends on the parameter $\kappa$. The FRG integrates out fluctuations of increasing rarity until it recovers the full EA $\Gamma$. In Fig.~\ref{fig:fRG_flow_schematic_EEOM} we show how this procedure gives us the EEOM. If one starts with the Langevin equation (\ref{eq:Langevin_sigma_dimless}) in the bottom left, we say that this can be described by some classical action (\ref{eq: PDF action_Sigma}) which we identify the REA with at some cutoff scale $\kappa = \Lambda \sim 1/\Delta \alpha$. This cutoff corresponds to fluctuations that occur over some timescale $\mathcal{O}(\Delta \alpha)$, if one wanted to simulate the Langevin equation (\ref{eq:Langevin_sigma_dimless}) this is the timestep they should use in their numerical scheme. The FRG then moves across the top line of Fig.~\ref{fig:fRG_flow_schematic_EEOM} integrating out fluctuations that occur over ever-increasing timescales until they are all integrated over and $\Gamma_{{\kappa}=0}$ is reached. One can then use (\ref{eq:gen QEOM}) to obtain the EEOM for the average position (\ref{eq:EEOM_sigma}) bringing us to the bottom right of Fig.~\ref{fig:fRG_flow_schematic_EEOM}. The flow equation (\ref{eq:dVsigma/dktilde2}) adapted from \cite{Synatschke2009,Wilkins2021} is shown at the bottom of Fig.~\ref{fig:fRG_flow_schematic_EEOM} as a straightforward way of moving between the Langevin equation (\ref{eq:Langevin_sigma_dimless}) and the EEOM (\ref{eq:EEOM_sigma}). In this way it is clear that the effective potential is the result of incorporating the fluctuating degrees of freedom hidden in the noise term $\eta$.\\
To obtain the effective potential one must consider a truncation of the derivative expansion. The leading order and next-to-leading order truncations are the Local Potential Approximation (LPA) and Wavefunction Renormalisation (WFR). The LPA assumes that the only object that changes as $\kappa$ is varied is the potential ($\bar{U} \rightarrow \bar{U}_{\kappa}$), whereas WFR also modifies the kinetic term in the action like so $\partial_{\alpha}\Sigma \rightarrow \zeta_{,\Sigma}\partial_{\alpha}\Sigma $. Here $\zeta$ depends on $\kappa$ and the subscript with a comma indicates a partial derivative. We are justified in a derivative expansion as we are dealing with overdamped motion where higher order time derivatives are slow-roll suppressed. Therefore, in some sense, higher order terms in the derivative expansion correspond to deviations from exact slow-roll that are induced by the stochastic fluctuations\footnote{The authors thank the anonymous referee for identifying this link.}.  Noting that the REA must match the classical action (\ref{eq: PDF action_Sigma}) at $\kappa = \Lambda$ the LPA form of the REA is:
\begin{eqnarray}
\Gamma_{\kappa}[\Sigma,\tilde{\Sigma},\tilde{C},C] = \int \text{d}\alpha~\bigg[\dfrac{\hat{H}^2}{2}\tilde{\Sigma}^2 - i\tilde{\Sigma}(\partial_{\alpha}\Sigma+ \hat{U}_{\kappa}') - \bar{C}\left( \partial_{\alpha} + \hat{U}_{\kappa}'' \right)C \bigg]\label{LPA Gamma}
\end{eqnarray}
where $\hat{U}_{\kappa = \Lambda} = \hat{U}$. The WFR functional form is more complicated and rather cumbersome. For completeness we reproduce equation (94) of \cite{Wilkins2021} here, such that the (on-shell)\footnote{Here this means that we integrated out the $\tilde{\Sigma}$ field for the sake of brevity.} effective action assuming WFR can be written as:
\begin{eqnarray}
	\Gamma_{\kappa}[\Sigma,\bar{C},C] = \int d\alpha~\dfrac{1}{2\hat{H}^2}\zeta_{,\Sigma}^{2}\lb\partial_{\alpha}{\Sigma}\rb^{2} + \dfrac{1}{2\hat{H}^2}\left(\dfrac{\hat{U}_{\kappa}'}{\zeta_{,\Sigma}}\right)^2 \nonumber \\
 - \bar{C}\left(\zeta_{,\Sigma}^2 \partial_{\alpha} + \zeta_{,\Sigma} \zeta_{,\Sigma\Sigma}\partial_{\alpha}{\Sigma} - \zeta_{,\Sigma\Sigma} \dfrac{\hat{U}_{\kappa}'}{\zeta_{,\Sigma}} + \hat{U}_{\kappa}''\right)C
\end{eqnarray}

\begin{figure}[t!]
    \centering
    \includegraphics[width = 0.45\linewidth]{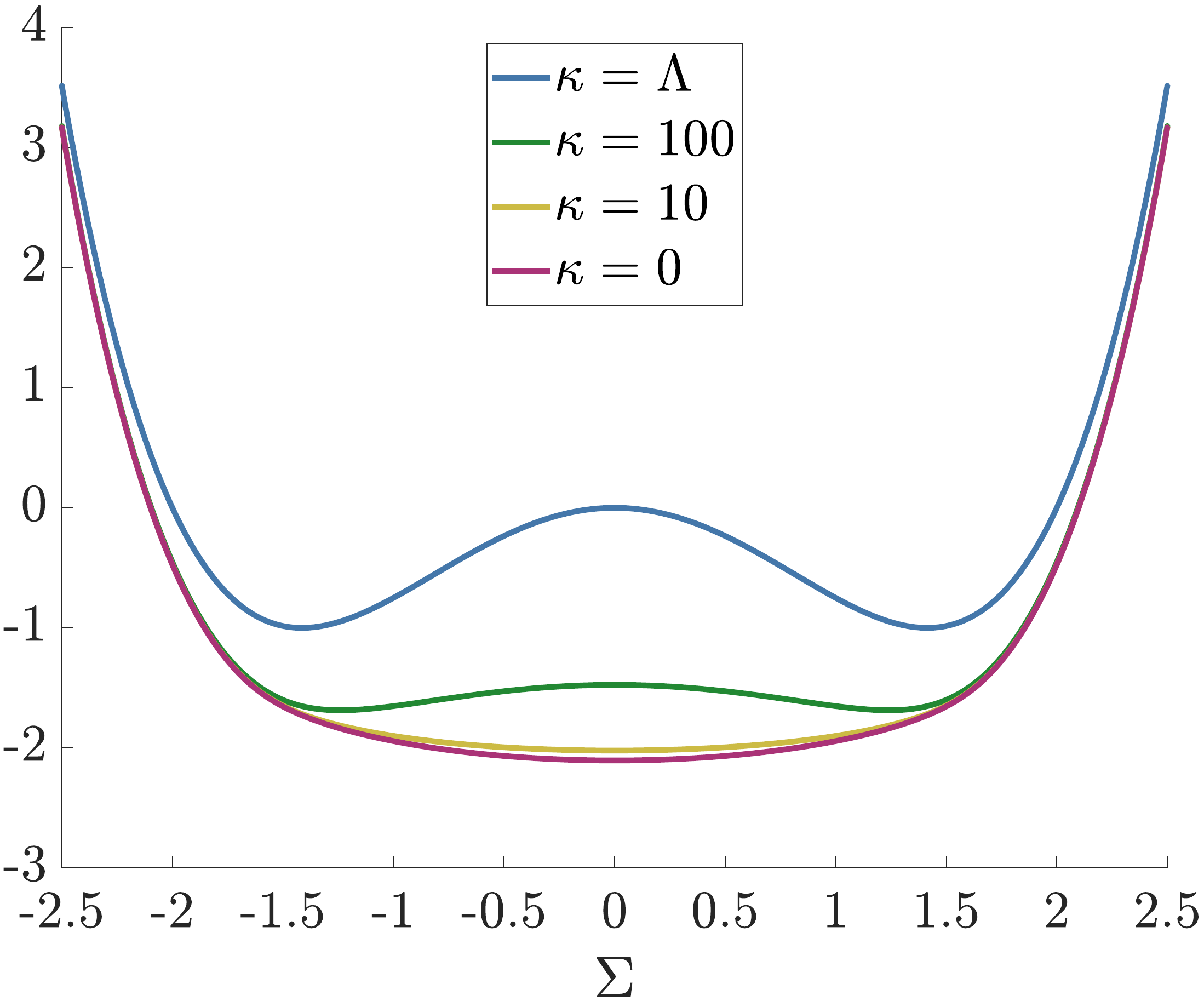}
    \includegraphics[width = 0.45\linewidth]{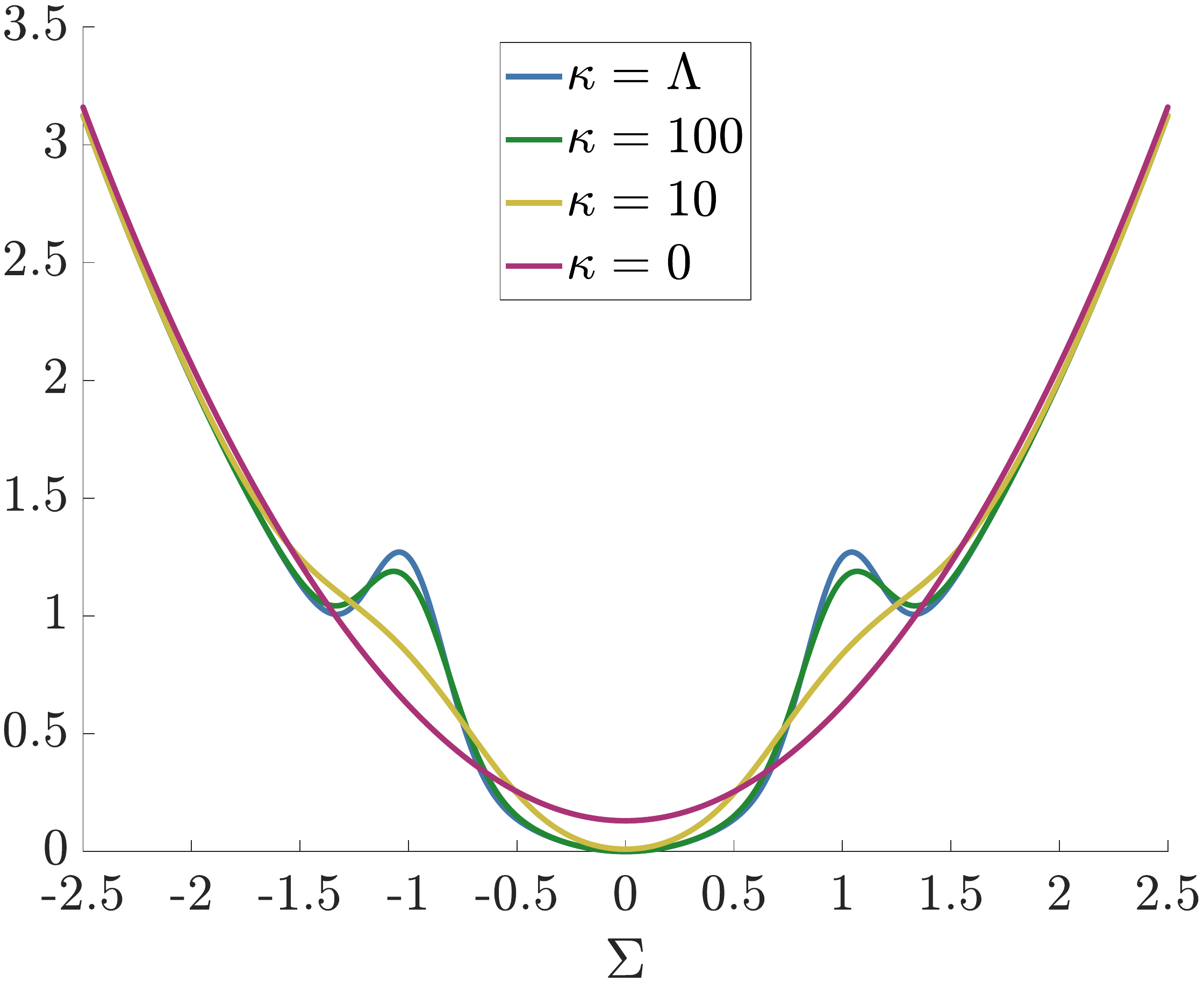}
    \caption{How the potential $\hat{U}_{\kappa}$ varies with renormalisation scale $\kappa$ for the doublewell potential (left) and the $\sigma^2$ plus 2 bumps potential (right) at $\hat{H}^2 = 2$. The cutoff in both cases has been chosen so as to correspond to a timestep of 0.01 i.e. $\Lambda = 2\pi \times 100$.}
    \label{fig:potential_flow}
\end{figure}

Under the LPA and WFR assumptions the flow equations for $\bar{U}_{\kappa}$ and $\zeta_{,\Sigma}$ are:
\begin{eqnarray}\label{eq:dVsigma/dktilde2}
\partial_{{{\kappa}}}\bar{U}_{{{\kappa}}}(\Sigma) &=& \dfrac{3\hat{H}^4}{4}\cdot\dfrac{1}{{{\kappa}}+ \partial_{\Sigma \Sigma}\bar{U}_{{{\kappa}}}(\Sigma)} \\
\partial_{{{\kappa}}}\zeta_{,\Sigma} &=& \dfrac{3\hat{H}^6}{2}\cdot\dfrac{\mathcal{P}}{\zeta_{,\Sigma}\cdot\mathcal{D}^2 } \label{eq:dzetax_sigma/dktilde}\\
\mathcal{D} &\equiv & \bar{U}_{,\Sigma\Sigma} + {\kappa}\,\zeta_{,\Sigma}^{2}, \quad \mathcal{P} \equiv \dfrac{4\zeta_{,\Sigma\Sigma} \bar{U}_{,\Sigma\Sigma\Sigma}}{\mathcal{D}} - \left( \zeta_{,\Sigma\Sigma}\zeta_{\Sigma}\right)_{,\Sigma} - \dfrac{3\zeta_{,\Sigma}^{2}\bar{U}_{,\Sigma\Sigma\Sigma}^{2}}{4\mathcal{D}^2} \nonumber \\
\end{eqnarray}
where we have rescaled the potential like so:
\begin{eqnarray}
\bar{U}(\Sigma ) = 3\hat{H}^2\hat{U} (\Sigma) \label{eq:Ubar_relation}
\end{eqnarray}
and $\kappa \in [0, ~ 3\hat{H}^2\Lambda]$. Equations (\ref{eq:dVsigma/dktilde2}) and (\ref{eq:dzetax_sigma/dktilde}) can be solved as outlined in \cite{Wilkins2021} and we will do so for a few different potentials. Namely the $\sigma^2$ plus 2 bumps, the doublewell and polynomial potentials:
\begin{eqnarray}
&& \sigma^2 + 2\text{ bumps: }  \bar{U}(\sigma) = \sigma^2 + \dfrac{3}{2}\lcb \exp \lsb-\dfrac{\lb \sigma - 1\rb^2}{0.06} \rsb + \exp \lsb -\dfrac{\lb \sigma + 1\rb^2}{0.06} \rsb \rcb \label{eq:varsigma^2_2bumpsdefn}\\
&& \text{ Doublewell: }   \bar{U}(\sigma) = -\sigma^2 + \dfrac{\sigma^4}{4}   \label{eq:sigma_DW_defn} \\
&& \text{ Polynomial: }   \bar{U}(\sigma) = \sigma + \dfrac{\sigma^2}{2} + \dfrac{2\sigma^3}{3} + \dfrac{\sigma^4}{4}   \label{eq:sigma_poly_defn}
\end{eqnarray}
In Fig.~\ref{fig:potential_flow} we plot how applying the LPA flow equation (\ref{eq:dVsigma/dktilde2}) modifies the potentials as $\kappa \rightarrow 0$ for the first two potentials of interest. We can see in both cases that as $\kappa \rightarrow 0$ the barriers disappear leaving a smooth, convex potential. For a more detailed analysis of the behaviour of the flow equations on various potentials see \cite{Wilkins2021}.
\subsection{One-point function}
The one-point function is simply the behaviour of the average field value $\Sigma (\alpha)= \lan \sigma (\alpha)\ran$. The late time, equilibrium value $\Sigma_{eq}$ is given by the minimum of the effective potential $\bar{U}_{{\kappa}=0}$. Given that it can take many e-folds for the system to relax to the de Sitter equilibrium, $P_{eq} = e^{-2\bar{U}/3\hat{H}^4}$, it is worth examining the non-equilibrium behaviour of the system at hand. We can determine the evolution of the average field value $\Sigma$ by a simple first order differential equation \cite{Wilkins2021}:
\begin{eqnarray}
\partial_{\alpha}{\Sigma} = -\tilde{U}_{,\Sigma}(\Sigma) \label{eq:EEOM_sigma}
\end{eqnarray} 
where we have introduced the \textit{effective dynamical potential} $\tilde{U}$ defined by 
\begin{eqnarray}
\tilde{U}_{,\Sigma}(\Sigma) \equiv 
\begin{cases}
\dfrac{\bar{U}_{,\Sigma}({\kappa} = 0, \Sigma)}{3\hat{H}^2}, & \text{ for LPA}  \\[10pt]
\dfrac{\bar{U}_{,\Sigma}({\kappa} = 0, \Sigma)}{3\hat{H}^2\zeta_{,\Sigma}^{2}({\kappa} = 0, \Sigma)}, & \text{ for WFR} 
\end{cases}\label{eq: Vtilde_sigma}
\end{eqnarray}
In \cite{Wilkins2021} it was shown that the FRG can very accurately predict the evolution of the one-point function both in and out of equilibrium.
\subsection{\label{sec:two-point}Two-point function}
\begin{figure}[t!]
    \centering
    \includegraphics[width = 0.45\linewidth]{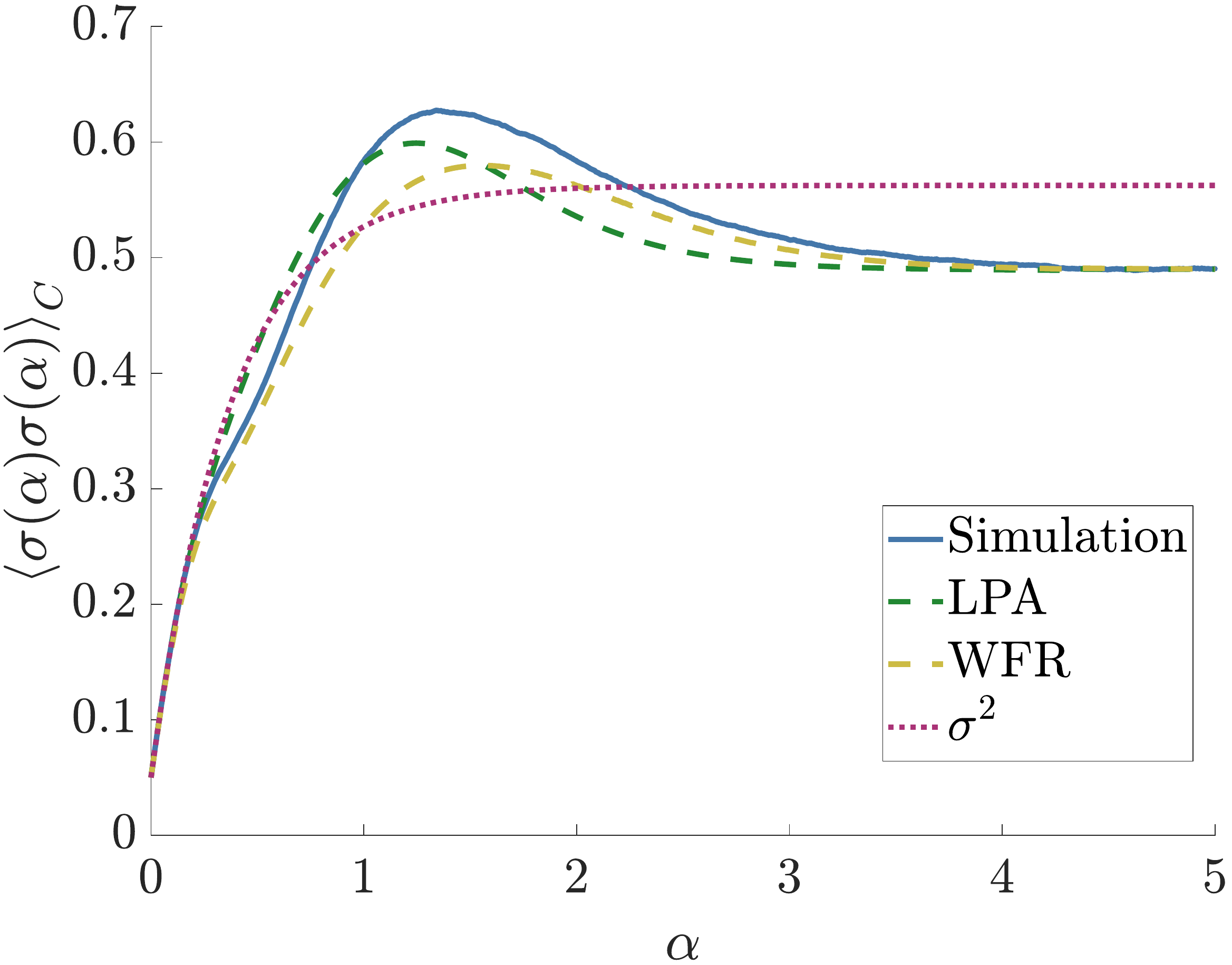}
    \includegraphics[width = 0.45\linewidth]{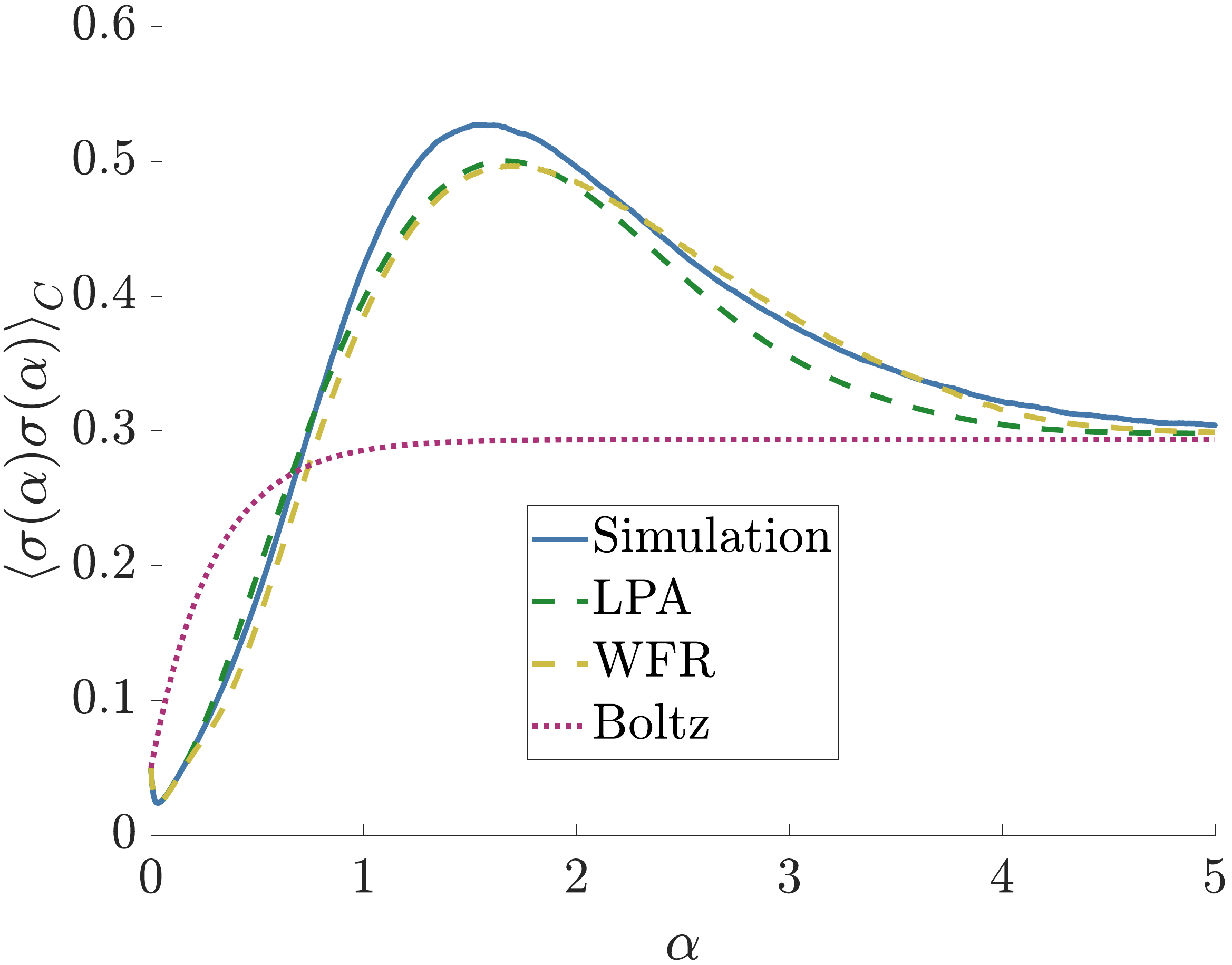}    
    \caption[Evolution of the variance for the polynomial and quadratic potential with bumps]{Evolution of the variance $\lan \sigma (\alpha)^2\ran_{C}$ in a $\sigma^2$ plus two bumps potential for $\hat{H}^2 = 1.5$ (left) and the polynomial potential at $\hat{H}^2 = 1$ (right) as computed by direct numerical simulation and from the FRG EEOM (\ref{eq:EEOM Variance}). The initial condition for both was a normal distribution with variance of 0.05 centred at $\sigma =3$.}
    \label{fig:Two_point}
\end{figure}
In equilibrium the connected two point function is straightforwardly modified from \cite{Wilkins2021} so that the connected correlation function at equilibrium is  
\begin{eqnarray}
\textbf{Cov}_{eq}(\sigma(\alpha_1)\sigma(\alpha_2)) = G_{eq}(\alpha_1, \alpha_2)  &=& \dfrac{3\hat{H}^4}{2\bar{U}_{,\Sigma\Sigma}\vert} e^{-\lambda|\alpha_1-\alpha_2|} \label{eq:2-pointfunc_sigma} \\
\textbf{Var}_{eq}(\sigma (\alpha)) = G_{eq}(\alpha, \alpha) &=& \dfrac{3\hat{H}^4}{2\bar{U}_{,\Sigma\Sigma}\vert}
\label{eq:equal2pt_sigma}
\end{eqnarray}
where $\lambda$ is given by 
\begin{eqnarray}\label{eq:lambdadef_sigma}
\lambda^2 \equiv  
\begin{cases}
\dfrac{\bar{U}_{,\Sigma\Sigma}^{2}\vert}{9\hat{H}^4}, & \text{for LPA}\\[10pt]
\dfrac{\bar{U}_{,\Sigma\Sigma}^{2}\vert}{9\hat{H}^4\zeta_{,\Sigma}^{4}\vert}, & \text{for WFR}
\end{cases} 
\end{eqnarray}
and the vertical line means the quantity should be evaluated at $\kappa = 0$ and $\Sigma = \Sigma_{eq}$. Note how the non-connected correlators follow the same behaviour:
\begin{eqnarray}
\lan \sigma (\alpha_1) \sigma ( \alpha _2) \ran_{eq} &=& G_{eq}(\alpha_1 , \alpha_2) + \lan \sigma (\alpha_1)\ran_{eq} \lan \sigma (\alpha_2)\ran_{eq} \\
\Rightarrow \lan \sigma (\alpha_1) \sigma ( \alpha _2) \ran _{eq} &=& \dfrac{3\hat{H}^4}{2\bar{U}_{,\Sigma\Sigma}\vert}  e^{-\lambda|\alpha_1-\alpha_2|} + \Sigma_{eq}^2 \label{eq:2pointfunc_equi_spect_dimless}
\end{eqnarray}
If we rearrange (\ref{eq:lambdadef_sigma}) so that\footnote{N.B. for LPA $\zeta_{,\Sigma}^{2}$ is unity.} $\bar{U}_{,\Sigma\Sigma} = 3\hat{H}^2\lambda \zeta_{,\Sigma}^{2}\vert $ and then restore (\ref{eq:2pointfunc_equi_spect_dimless}) to the true physical parameters we obtain:
\begin{eqnarray}
\lan \sigma (\alpha_1) \sigma ( \alpha _2) \ran _{eq} &=&  \dfrac{H^2}{8\pi^2\lambda\zeta_{,\hat{\Sigma}}\vert}  e^{-\lambda|\alpha_1-\alpha_2|}+ \Sigma_{eq}^2 \label{eq:2pointfunc_equi_spect_dimfull}
\end{eqnarray}
If however we assume the system has not yet reached equilibrium then the variance and covariance can be expressed more generally \cite{Wilkins2021}:
\begin{eqnarray}
\lan \sigma (\alpha) \sigma ( \alpha ) \ran_C &\equiv &G(\alpha,\alpha) = \dfrac{\hat{H}^2}{2\lambda P(\alpha)} \tilde{Y}_1(\alpha)\tilde{Y}_2(\alpha)  +~ \dfrac{P(0)}{P(\alpha)}\left[G_{00} - \dfrac{\hat{H}^2}{2\lambda P(0)}\right]\tilde{Y}_{2}^{2}(\alpha)
 \label{eq:EEOM Variance}\\
 \lan \sigma (\alpha_1) \sigma ( \alpha _2) \ran_C &\equiv &G(\alpha_1,\alpha_2) = G_{\alpha_1\alpha_1}\tilde{Y}_2(\alpha) \label{eq:EEOM Covariance}
\end{eqnarray}
which is now written in terms of the \emph{dimensionless} parameters. The functions $\tilde{Y}_1 (\alpha)$, $\tilde{Y}_2(\alpha)$ are the normalised solutions to:
\begin{eqnarray}
&& \quad \quad \quad \quad \quad \ddot{f}(\alpha) -  \mathcal{U}(\Sigma (\alpha))f(\alpha) = 0 \label{eq:appendhomo2ptfunc} \\
\mathcal{U}(\Sigma) &=& 
\begin{cases}
    \bar{U}_{,\Sigma\Sigma}^{2} + \bar{U}_{,\Sigma}\bar{U}_{,\Sigma\Sigma\Sigma}, & \text{for LPA}\\[10pt]
   % \dfrac{\bar{U}_{,\Sigma\Sigma}^{2}}{\zeta_{,\Sigma}^{4}}+ \dfrac{\bar{U}_{,\Sigma}\bar{U}_{,\Sigma\Sigma\Sigma}}{\zeta_{,\Sigma}^{4}}- \dfrac{\bar{U}_{,\Sigma}^{2}\zeta_{,\Sigma\Sigma\Sigma}}{\zeta_{,\Sigma}^{5}}  \\
 %-\dfrac{5\bar{U}_{,\Sigma}\bar{U}_{,\Sigma\Sigma}\zeta_{,\Sigma\Sigma}}{\zeta_{,\Sigma}^{5}}    + \dfrac{5\bar{U}_{,\Sigma}^{2}\zeta_{,\Sigma\Sigma}^{2}}{\zeta_{,\Sigma}^{6}} , & \text{for \ac{WFR}}
 \dfrac{\bar{U}_{,\Sigma\Sigma}^{2}}{\zeta_{,\Sigma}^{4}}+ \dfrac{\bar{U}_{,\Sigma}\bar{U}_{,\Sigma\Sigma\Sigma}}{\zeta_{,\Sigma}^{4}} -\dfrac{4\bar{U}_{,\Sigma}\bar{U}_{,\Sigma\Sigma}\zeta_{,\Sigma\Sigma}}{\zeta_{,\Sigma}^{5}}    + \dfrac{4\bar{U}_{,\Sigma}^{2}\zeta_{,\Sigma\Sigma}^{2}}{\zeta_{,\Sigma}^{6}} , & \text{for WFR}
  \end{cases} \label{eq:U = }
\end{eqnarray}
and $P(\alpha) = 1$ or $\zeta_{\Sigma}^2(\Sigma(\alpha))$ for LPA and WFR respectively. In \cite{Wilkins2021} it was shown that the FRG can very accurately predict the evolution of the two-point function both in and out of equilibrium although generically not as well as it predicts the one-point function. In Fig.~\ref{fig:Two_point} we plot the evolution of the variance in a $\sigma^2$ plus two bumps potential and the polynomial potential as calculated by direct numerical simulation and the FRG. We have also included, as a baseline comparison, the evolution in a bare $\sigma^2$ potential and the Boltzmann potential -- see discussion around (\ref{eq: BoltztildeU}). As expected from \cite{Wilkins2021}, the FRG offers good agreement with direct numerical simulation and is a dramatic improvement over alternative, naive estimates. 

\subsection{Three-point function}
Only the FRG predictions for the one-point and connected two-point function were derived in \cite{Wilkins2021}. Here we will extend this analysis to the three point function by use of the formula -- see e.g. pages 381-382 \cite{Peskin:1995ev}:
\begin{eqnarray}
\lan \sigma_a \sigma_b \sigma_c \ran_{C} = \int \mathrm{d}u\mathrm{d}v\mathrm{d}w  ~\dfrac{\delta^3\Gamma [\Sigma]}{\delta \Sigma_u \delta \Sigma_v\delta \Sigma_w}~G_{au}G_{bv}G_{cw}
\end{eqnarray}
where subscripts indicate the argument and $G_{ab} \equiv \lan \sigma(a) \sigma(b) \ran_{C}$. The third functional derivative of the EA can be computed as:
\begin{eqnarray}
\dfrac{\delta^3\Gamma [\Sigma]}{\delta \Sigma_u \delta \Sigma_v\delta \Sigma_w} &=& \dfrac{1}{\hat{H}^2}\lsb 2\zeta_{,\Sigma}\zeta_{,\Sigma\Sigma} \partial_{ww} - \mathcal{W}(\Sigma)\rsb \delta (w-u) \delta (w-v) \\
\mathcal{W}(\Sigma) &\equiv & 3\dfrac{\hat{U}_{,\Sigma\Sigma}\hat{U}_{,\Sigma\Sigma\Sigma}}{\zeta_{,\Sigma}^2} + \dfrac{\hat{U}_{,\Sigma}\hat{U}_{,\Sigma\Sigma\Sigma\Sigma}}{\zeta_{,\Sigma}^2} - 6\dfrac{\hat{U}_{,\Sigma\Sigma}^2 \zeta_{,\Sigma\Sigma}}{\zeta_{,\Sigma}^3} \nonumber \\
&& -6\dfrac{\hat{U}_{,\Sigma}\hat{U}_{,\Sigma\Sigma\Sigma}\zeta_{,\Sigma\Sigma}}{\zeta_{,\Sigma}^3} -9\dfrac{\hat{U}_{,\Sigma}\hat{U}_{,\Sigma\Sigma}\zeta_{,\Sigma\Sigma\Sigma}}{\zeta_{,\Sigma}^3} -2\dfrac{\hat{U}_{,\Sigma}^2\zeta_{,\Sigma\Sigma\Sigma\Sigma}}{\zeta_{,\Sigma}^3} \nonumber \\
&&+18\dfrac{\hat{U}_{,\Sigma}\hat{U}_{,\Sigma\Sigma}\zeta_{,\Sigma\Sigma}^2}{\zeta_{,\Sigma}^4} + 9 \dfrac{\hat{U}_{,\Sigma}^2 \zeta_{,\Sigma\Sigma}\zeta_{,\Sigma\Sigma\Sigma}}{\zeta_{,\Sigma}^4} -12\dfrac{\hat{U}_{,\Sigma}^2\zeta_{,\Sigma\Sigma}^3}{\zeta_{,\Sigma}^5}
\end{eqnarray}
If we modify appropriately the initial conditions we can write the connected two point function like so:
\begin{eqnarray}
G_{ab} = \Theta (b-a) G_{aa}\dfrac{G_{0b}}{G_{0a}}
\end{eqnarray}
which we can combine to give the following EEOM for the third central moment:
\begin{eqnarray}
\lan \sigma_{\alpha}^3\ran_{C} = \lb\dfrac{G_{\alpha\alpha}}{G_{0\alpha}}\rb^3 \int_{\alpha}^{\infty}\mathrm{d}\tilde{\alpha}~ \Bigg\{ \dfrac{6\zeta_{\Sigma}\zeta_{\Sigma\Sigma}}{\hat{H}^2}\lsb \lb G_{0\tilde{\alpha}}\rb^2 \partial_{\tilde{\alpha}\tilde{\alpha}}G_{0\tilde{\alpha}} + 2G_{0\tilde{\alpha}}\lb \partial_{\tilde{\alpha}}G_{0\tilde{\alpha}}\rb^2 \rsb  - \dfrac{\lb G_{0\tilde{\alpha}}\rb^3}{\hat{H}^2}\mathcal{W}(\Sigma (\tilde{\alpha}))\Bigg\} \nonumber \\ \label{eq:FRG_threepoint}
\end{eqnarray}
The equilibrium  limit is much simpler and assuming that $\alpha_3 \geq \alpha_2 \geq \alpha_1$ can be written as:
\begin{eqnarray}
\left\langle \sigma(\alpha_1)\sigma(\alpha_2)\sigma(\alpha_3) \right\rangle_{C}  &=& \left\langle \sigma(\alpha_1)^3 \right\rangle_{C} e^{-\lambda (2\alpha_3 -\alpha_2-\alpha_1)} \label{3-pointfunc}\\
\left\langle \sigma(\alpha_1)^3 \right\rangle_{C} & = & -\left\langle \sigma(\alpha_1)^2 \right\rangle_{C}^{3}\dfrac{\mathcal{W}(\Sigma_{eq})}{3\lambda} \label{eq:three-point_equi}
\end{eqnarray}
with the potential $\mathcal{W}$ evaluated at the equilibrium point $\Sigma_{eq}$. In Table \ref{tabel:Equi_Threept} we write down the FRG prediction for the equilibrium third central moment in the polynomial potential for a few values of $\hat{H}^2$ and compare it to the value from the Boltzmann distribution. We can see that there is reasonable agreement for $\hat{H}^2 = 2$, $1.5$ with the LPA proving more accurate than WFR. However as $\hat{H}^2$ is increased/decreased we find the agreement is poorer. For the other two, symmetric, potentials the FRG predicts the correct zero value of the third central moment.\\
\begin{table}[t!]
	\centering
	\begin{tabular}{|l |l l l|}
		\hline
		$\hat{H}^2$ & Boltz & LPA & WFR \\
		\hline
		5 & 0.3787 & 0.9480& 0.7078\\
            2 & 0.3778 & 0.3858& 0.2932\\
            1.5 & 0.3236& 0.2510& 0.1988\\
            1 & 0.1547 & 0.08288& 0.07868\\
            0.5 & 0.005128 & 0.001287 & 0.001533\\  
		\hline 
	\end{tabular}\caption{\label{tabel:Equi_Threept}The equilibrium third central moment $\left\langle \sigma(\alpha_1)^3 \right\rangle_{C}$ as calculated from the Boltzmann equilbtium distribution $P_{eq} = e^{-2\bar{U}/3\hat{H}^4}$, the LPA and WFR (\ref{eq:three-point_equi}) for the polynomial potential.} 
	
\end{table}
\begin{figure}[t!]
    \centering
    \includegraphics[width = 0.45\linewidth]{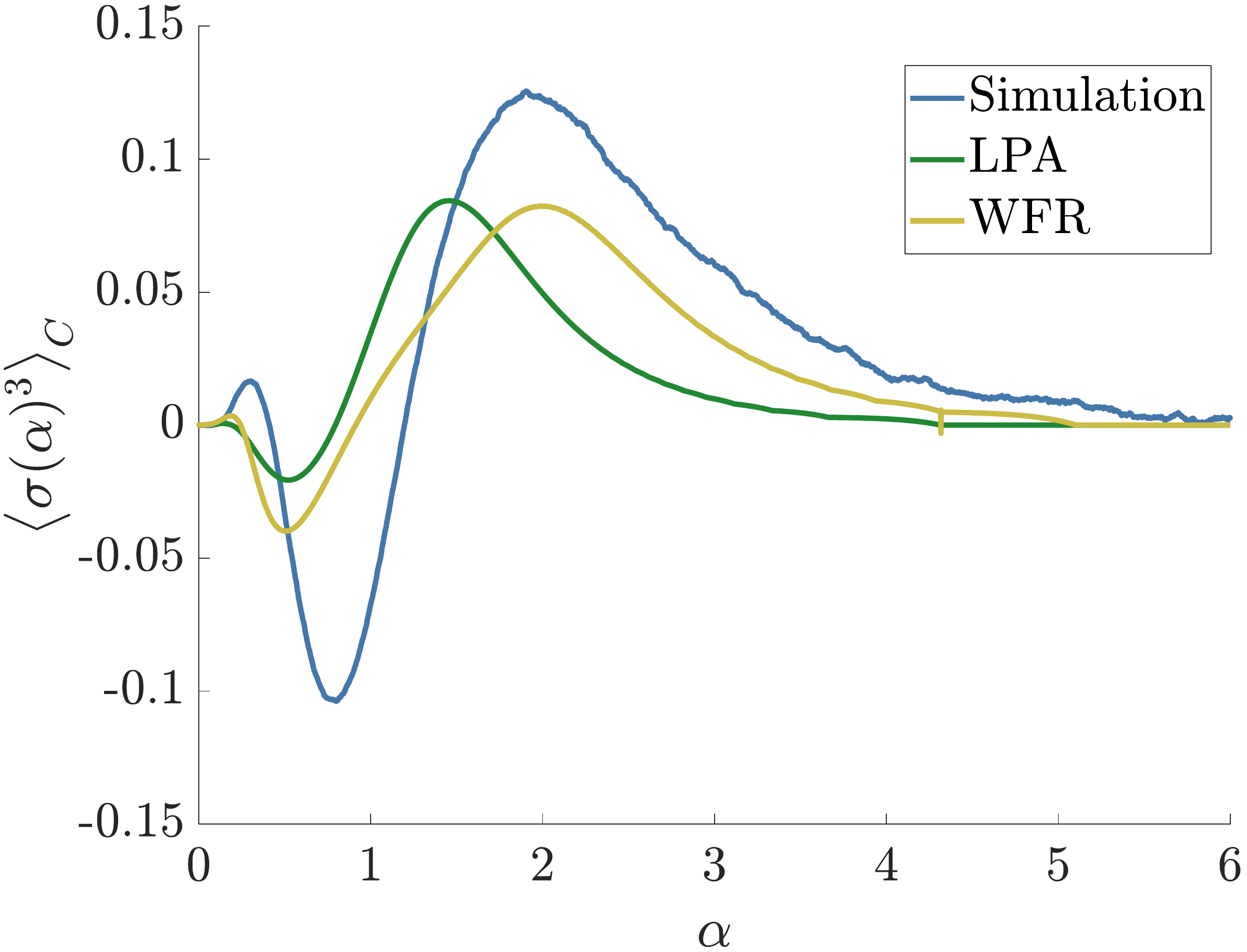}
    \includegraphics[width = 0.45\linewidth]{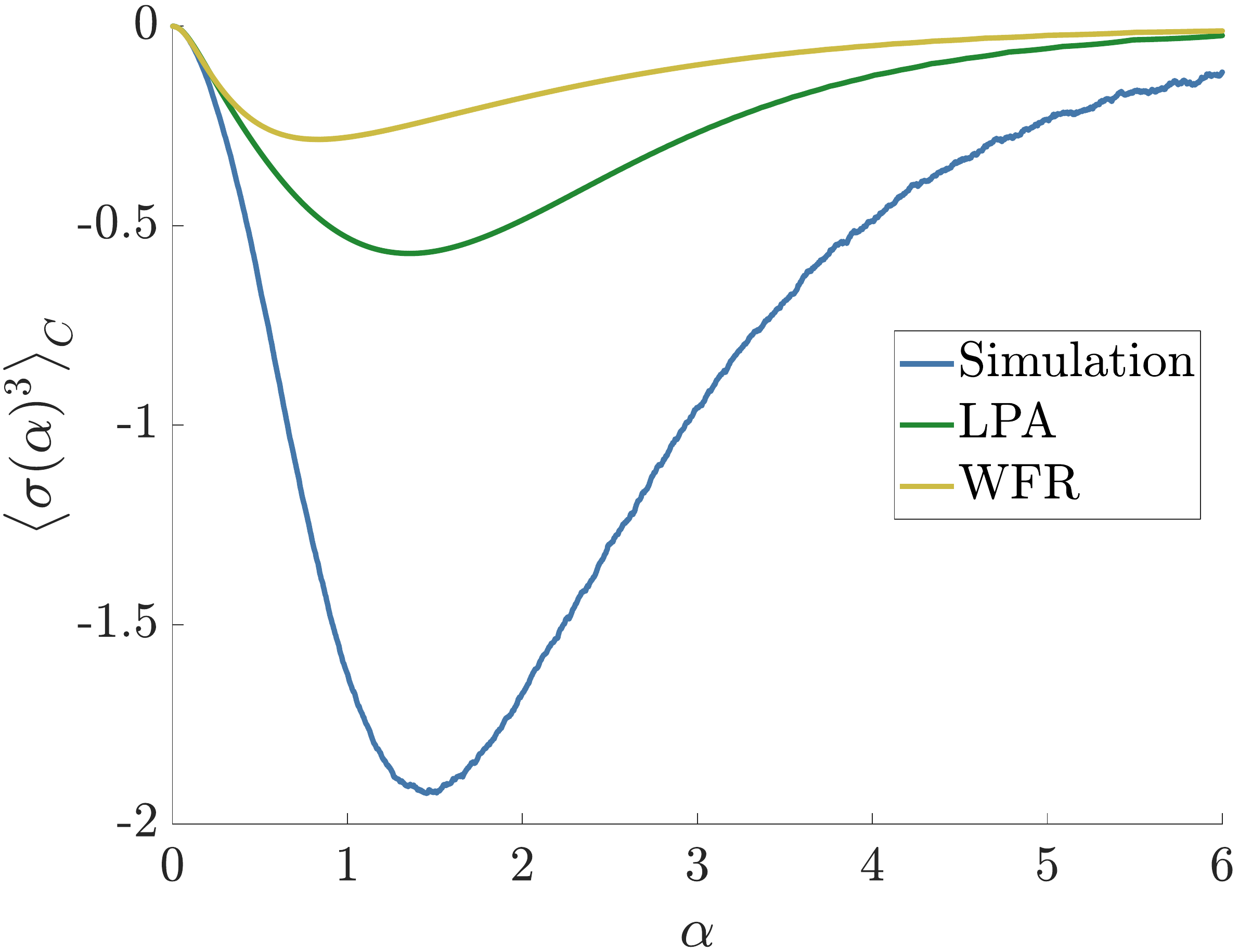}
    \caption[Evolution of the third central moment for the doublewell and quadratic potential with bumps]{Evolution of the third central moment $\lan \sigma (\alpha)^3\ran$ in a $\sigma^2$ plus two bumps potential for $\hat{H}^2 = 1.5$ (left) and the doublewell potential at $\hat{H}^2 = 5$ (right) as computed by direct numerical simulation and from the FRG EEOM (\ref{eq:FRG_threepoint}). The initial condition for both was a normal distribution with variance of 0.05 centred at $\sigma =3$.}
    \label{fig:Three_point}
\end{figure}
In Fig.~\ref{fig:Three_point} we therefore plot the solution to the full EEOM (\ref{eq:FRG_threepoint}) for the $\sigma^2$ plus two bumps potential (left plot) and doublewell (right plot) for favourable choices of $\hat{H}^2$ and compare to direct numerical simulations. It is clear that the FRG can capture the qualitative evolution of the third central moment reasonably well, however it is not very precise and we were unable to improve the accuracy for different choices of initial conditions or $\hat{H}^2$. It would therefore seem that we have reached the limit of reasonable accuracy that the current FRG procedure can provide. It is possible that one needs to go to higher order in the derivative expansion to get accurate results for the third central moment, or perhaps one should instead look at a vertex expansion \cite{Morris1994} of the FRG approach instead.  \\

%Where we have introduced the notion of the \textit{bosonic potential} $\mathcal{U}$:
%\begin{eqnarray}
%\mathcal{U}({\kappa},\Sigma) \equiv  
%\begin{cases}
%U_{,\Sigma}^{2}({\kappa},\Sigma), & \text{for LPA}\\[10pt]
%\dfrac{U_{,\Sigma}^{2}({\kappa},\Sigma)}{\zeta_{,\Sigma}^{2}({\kappa},\Sigma)}, & \text{for WFR}
%\end{cases}
%\label{eq:bosondef}
%\end{eqnarray}
%Similarly the connected 4-point function at equilibrium is:
%\begin{eqnarray}
%&&\left\langle \sigma(\alpha_1)\sigma(\alpha_2)\sigma(\alpha_3)\sigma(\alpha_4) \right\rangle_{C} = \left\langle \sigma(t_1)^4 \right\rangle_{C}e^{-\lambda (3\alpha_4 -\alpha_3 -\alpha_2-\alpha_1)} \nonumber \\
%\label{4-pointfunc}\\
%&&\left\langle \sigma(t_1)^4 \right\rangle_{C} =  \dfrac{\left\langle \sigma(\alpha_1)^2 \right\rangle_{C}^{4}}{4\lambda}\left(\mathcal{U}_{,\Sigma\Sigma\Sigma}^{2}\vert\dfrac{\left\langle \sigma(\alpha_1)^2 \right\rangle_{C}}{\lambda} - \mathcal{U}_{,\Sigma\Sigma\Sigma\Sigma}\vert\right) \nonumber \\
%\end{eqnarray}
%From these connected correlation functions one should be able to calculate the skewness and kurtosis of the equilibrium distribution in a similar manner to what we did for $\Sigma_{eq}$ and $\textbf{Var}(\sigma)$. 

\section{\label{sec:Cosmo_Spec} Cosmological Observables}

The correlators obtained via the stochastic dynamics are temporal and, by construction of the stochastic theory, apply to successive times within a single Hubble patch. However, because of de Sitter invariance \cite{Starobinsky1994} any correlator of a scalar observable $\mathcal{O}(\sigma)$ should only depend on the de Sitter invariant quantity:
\begin{eqnarray}
y = \text{cosh} \lb \alpha_1 - \alpha_2  \rb -\dfrac{H^2}{2}\exp \lb \alpha_1 + \alpha_2  \rb \vert \vec{r}_1 - \vec{r}_2\vert^2
\end{eqnarray}
where $\vec{r}_1$ and $\vec{r}_2$ are comoving position vectors. One might therefore expect that purely spatial correlators can be obtained from temporal ones by mapping the temporal interval to a spatial one with the same value of $\vert y \vert$. In practise, this can be done via the correspondence 
\begin{equation}
e^{\left(\alpha_1 - \alpha_2\right)} \leftrightarrow  H^2 \exp \lb 2\alpha_2  \rb \vert \vec{r}_1 - \vec{r}_2\vert^2
\end{equation}
where we assumed that $\alpha_1 - \alpha_2 >1$. It can be explicitly checked that this procedure gives the correct spatial dependence of the correlators in the case of a massive free field to leading order for large $\vert y\vert\gg 1$ - see e.g.~\cite{Garbrecht2014, Belgacem2022}. Furthermore, the diagrammatic correspondence on long wavelengths between the stochastic theory and the full QFT with quartic self-interactions shown in \cite{Garbrecht2015} suggests that this procedure would provide the correct spatial correlator in that case too. Here we will assume that this substitution can be done in general, as for example used in \cite{Markkanen2019} and write 
\begin{eqnarray}
\lan \sigma (\vec{x}_1,\alpha) \sigma (\vec{x}_2,\alpha)\ran = \lan \sigma(\alpha) \sigma  (\alpha + 2\ln \lb \left| \vec{x}_1 - \vec{x}_2\right| H\rb) \ran \label{eq:equal_time=equal_space}
\end{eqnarray}
where  the right hand side is the temporal correlation function at spatial coincidence $(\vec{r}_1 = \vec{r}_2)$ that can in principle be obtained through the stochastic approach i.e. equation (\ref{eq:Langevin_sigma}). This is valid at distances $\vert \vec{x}_1 -\vec{x}_2\vert \gg 1/H$ and $\vec{x} = a\vec{r}$ is the physical, non-comoving coordinate and $\sigma$ is now \emph{dimensionful}. Equal-time correlation functions are often described by their power spectrum:
\begin{eqnarray}
\mathcal{P}_{\sigma}(k) = \dfrac{k^3}{2\pi^2} \int \mathrm{d}^3x~e^{-i\Vec{k}\cdot \Vec{x} }\lan \sigma (\Vec{x}_1,\alpha) \sigma (\Vec{x}_2,\alpha)\ran \label{eq:power_spec_sigma_defn}
\end{eqnarray}
where here $k$ is the Fourier transform of position and $\Vec{x} = \Vec{x}_1 - \Vec{x}_2$. The question now is how the FRG can compute the RHS of (\ref{eq:equal_time=equal_space}) and therefore the power spectrum. 

\subsection{\label{sec:powerspec_equi} Power Spectrum from an equilibrium distribution}
In equilibrium the FRG predicts that the two point function follows a simple exponential decay (\ref{eq:2pointfunc_equi_spect_dimfull}) -- for simplicity we assume a symmetric potential such that $\Sigma_{eq} = 0$. We can substitute this into the RHS of (\ref{eq:equal_time=equal_space}) to obtain:
\begin{eqnarray}
\lan \sigma (\Vec{x}_1,\alpha) \sigma (\Vec{x}_2,\alpha)\ran = \dfrac{H^2}{8\pi^2\lambda\zeta_{,\hat{\Sigma}}\vert}  \dfrac{1}{\lb \left| \Vec{x}_1 - \Vec{x}_2\right| H\rb^{2\lambda}}
\end{eqnarray}
Which suggests a power law form:
\begin{eqnarray}
\lan \sigma (\Vec{x}_1,\alpha) \sigma (\Vec{x}_2,\alpha)\ran &=& \dfrac{A_{\sigma}}{\lb \left| \Vec{x}_1 - \Vec{x}_2\right| H\rb^{n_{\sigma} -1}} \\
A_{\sigma} &=& \dfrac{H^2}{4\pi^2}\dfrac{1}{\zeta_{,\hat{\Sigma}}\vert \lb n_{\sigma} -1\rb}  \label{eq:powerspec_ampli_sigma}\\
n_{\sigma} &=& 1 + 2\lambda \label{eq:spectral_tilt_sigma}
\end{eqnarray}
Using the definition of the power spectrum (\ref{eq:power_spec_sigma_defn}) we obtain\footnote{It is worth noting that this simple form assumes that $\vert n_{\sigma} -1 \vert \ll 1$ otherwise the power spectrum is more generally given by:
\begin{eqnarray}
\mathcal{P}_{\sigma}(k) &=& \dfrac{2}{\pi}A_{\sigma}\Gamma \lsb 2 - 2\lambda \rsb \sin \lb \pi\lambda\rb \lb\dfrac{k}{H}\rb^{n_{\sigma} -1} 
\end{eqnarray}}
\begin{eqnarray}
\mathcal{P}_{\sigma}(k) = A_{\sigma}(n_{\sigma} -1) \lb\dfrac{k}{H}\rb^{n_{\sigma} -1} 
\end{eqnarray}
which suggests that $A_{\sigma}$ and $n_{\sigma}$ are the amplitude of the power spectrum and the spectral tilt respectively for $\sigma$. In terms of FRG quantities the power spectrum is given by:
\begin{eqnarray}
\mathcal{P}_{\sigma}(k) =   \dfrac{H^2}{4\pi^2\zeta_{,\hat{\Sigma}}\vert}  \lb\dfrac{k}{H}\rb^{2\lambda}\label{eq:Power_spec_sig}
\end{eqnarray}
\begin{table}[t!]
	\centering
		\begin{tabular}{|l |l l l l|}
			\hline
			$\hat{H}^2$ & LPA & WFR & Sim & Bare\\
			\hline
			5 & 0.7902  & 0.7874 & 0.7841 & 0.8\\
			2 & 1.9916 & 1.8427 &  1.8703 & 2\\
			1 & 6.4109 & 5.6368 & 5.3684 & 4\\
			0.5 & 9.3241 & 9.2108 & 9.2849 & 8\\
			\hline
		\end{tabular}
	 \caption[Value of spectral tilt as computed by FRG and by direct simulation for a harmonic potential with Gaussian bumps.]{Value of the (shifted) spectral tilt $n_{\sigma} - 1$ as computed by the LPA, WFR and by direct numerical simulation for the $\sigma^2$ plus bumps potential. Increasing $\hat{H}^2$ increases the strength of the fluctuations. The simulation values were generated by averaging over 50,000 runs. In the final column we have included, for comparison, what the prediction from the underlying harmonic potential would be without the Gaussian bumps.}
	  	 \label{tabel:spectral_tilt}
\end{table}
\begin{figure}[t!]
    \centering
    \includegraphics[width = 0.45\linewidth]{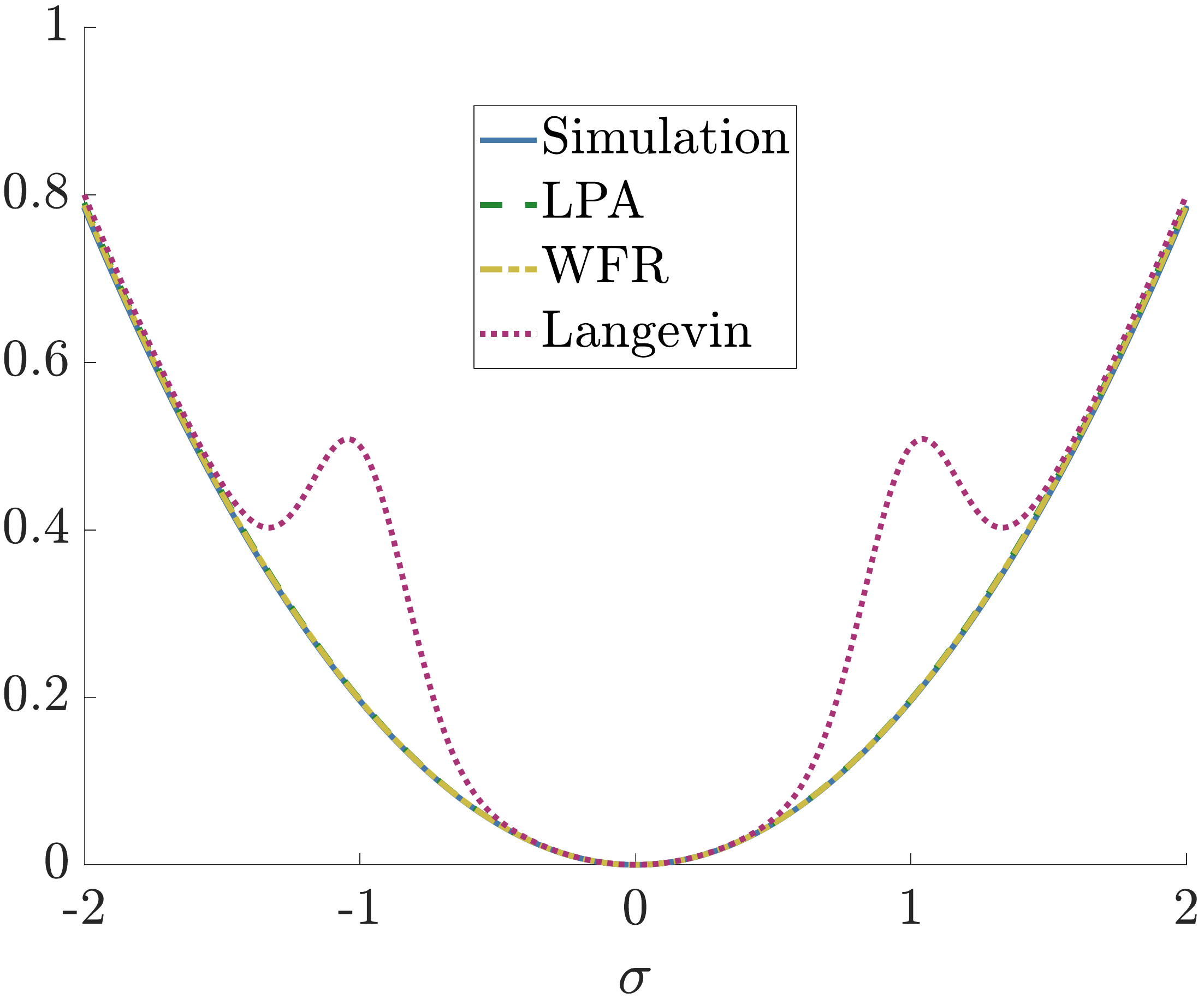}
    \includegraphics[width = 0.45\linewidth]{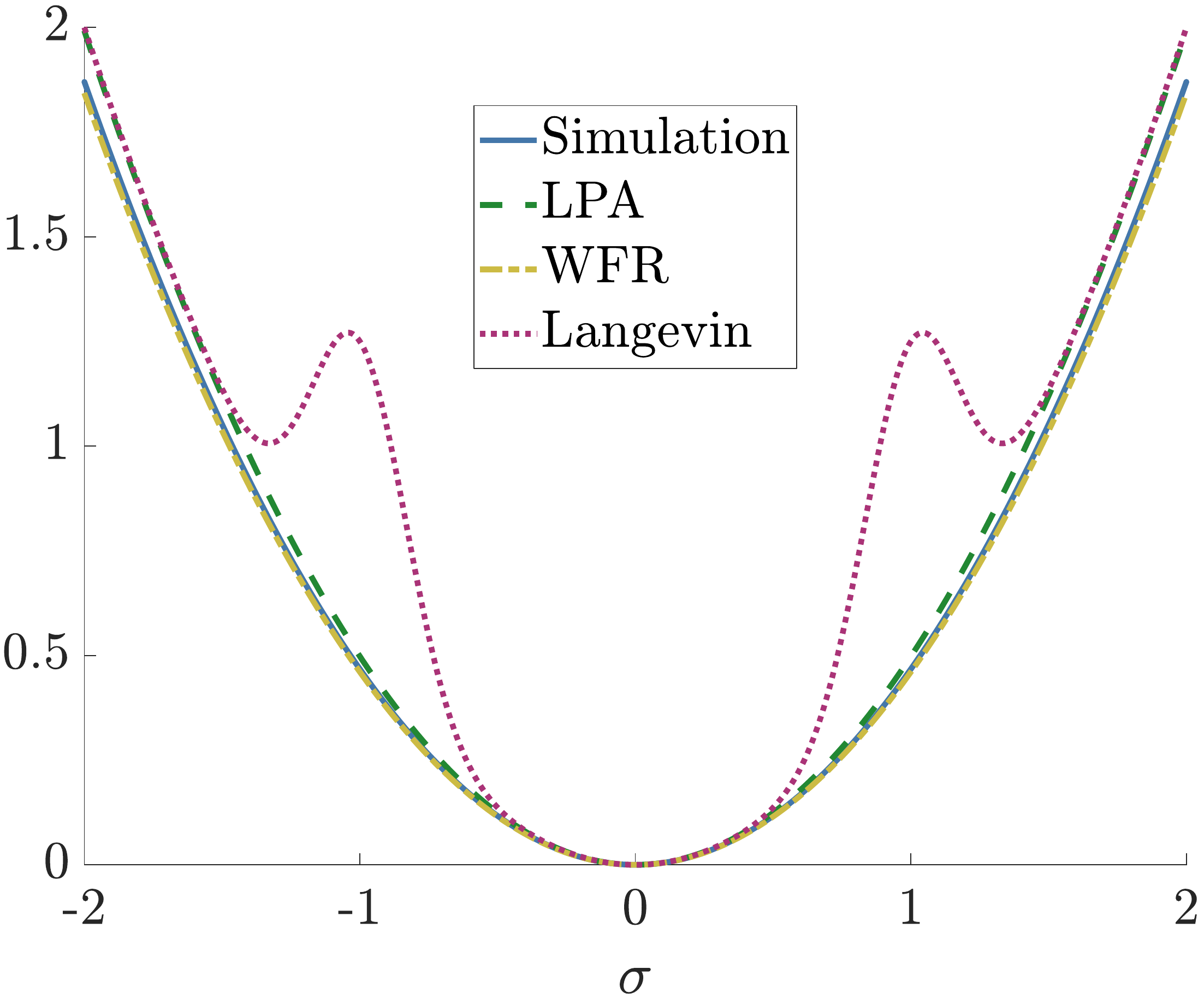}
    \includegraphics[width = 0.45\linewidth]{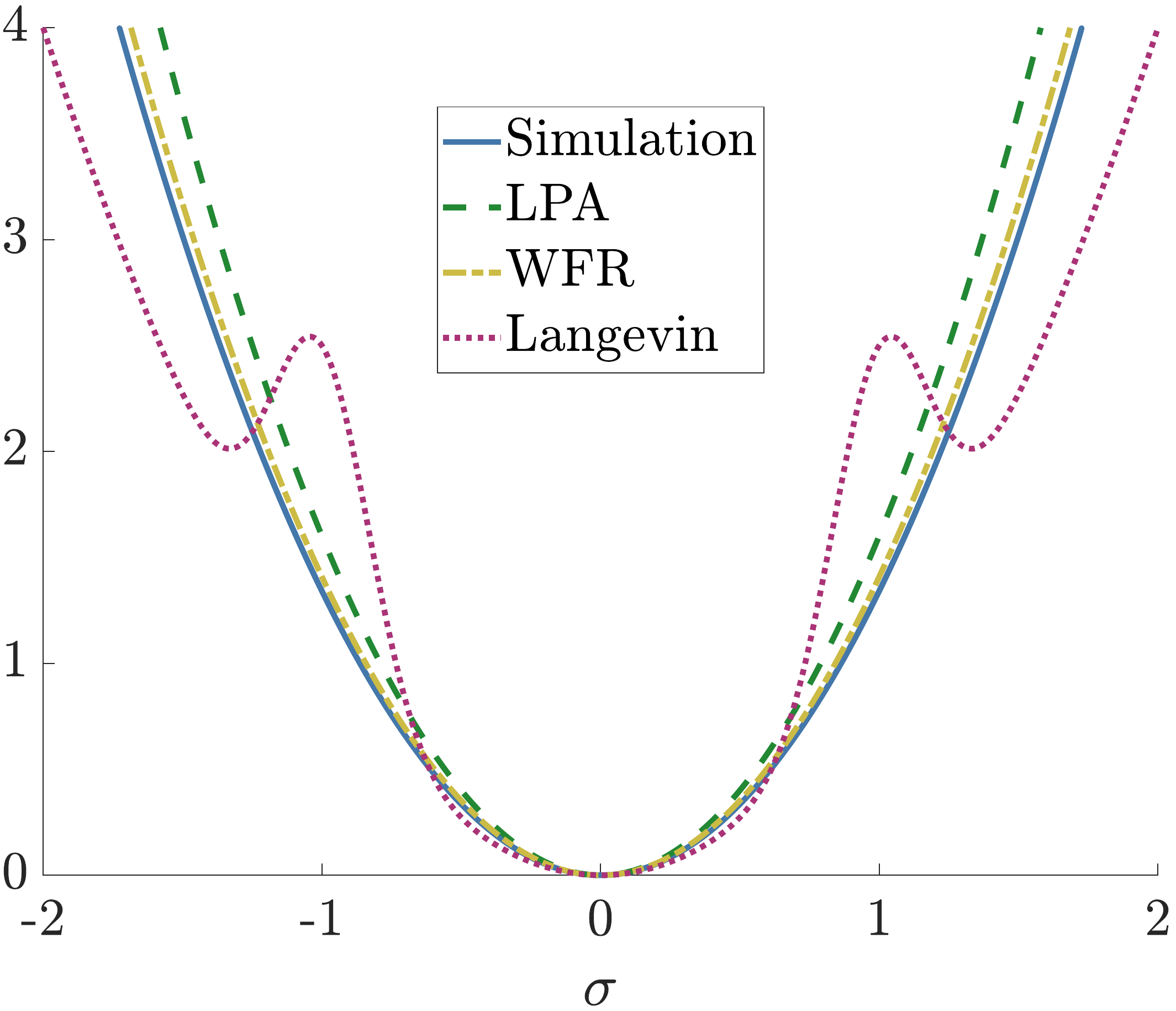}
    \includegraphics[width = 0.45\linewidth]{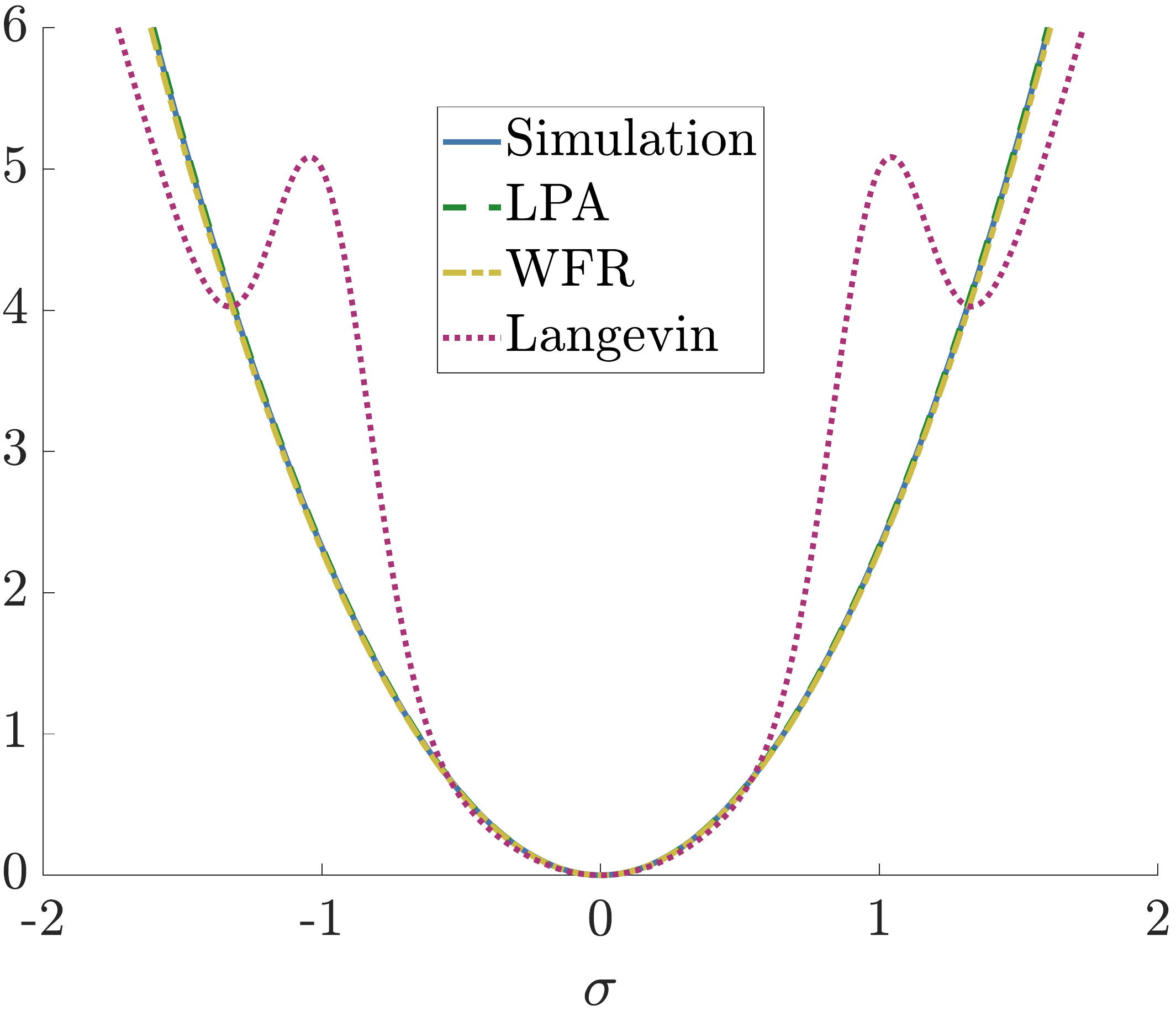}
    \caption[Different potentials predicting the same spectral tilt]{Various harmonic potentials $U(\sigma) \propto \sigma^2$ that give the same prediction for $\lambda$ and therefore the spectral tilt $n_{\sigma}$ as computed by direct simulation, the LPA and WFR for the $\sigma^2$ plus bumps potential -- shown here by the dotted red line. The parameters for each subplot are: $\hat{H}^2 = 5$ (top left), $\hat{H}^2 = 2$ (top right), $\hat{H}^2 = 1$ (bottom left)  and $\hat{H}^2 = 0.5$ (bottom right). }
    \label{fig:spectral_tilt_compare}
\end{figure}
 The parameter $\lambda$ defined in (\ref{eq:lambdadef_sigma}) therefore take on a new interpretation, it tell us how accurately the FRG can predict the power spectrum of a spectator field $\sigma$. In \cite{Markkanen2019,Markkanen2020} the stochastic spectral expansion is used to obtain the amplitude of the power spectrum and the spectral tilt for a standard fourth order polynomial and doublewell potential. Here we will use FRG techniques to compute the spectral tilt for the $\sigma^2$ plus two bumps potential (\ref{eq:varsigma^2_2bumpsdefn}) while varying $\hat{H^2}$. \\
 
It is straightforward to solve the appropriate flow equations to obtain $\lambda$ from the FRG and then using (\ref{eq:spectral_tilt_sigma}) obtain the (shifted) spectral tilt $n_{\sigma} -1$. In Table~\ref{tabel:spectral_tilt} we compare these computations to the result from direct numerical simulation as well as what the spectral tilt would be for the simple underlying harmonic potential in the absence of Gaussian bumps. In line with the results of \cite{Wilkins2021} we can see good agreement using FRG techniques with WFR offering improvement over the LPA result. As indicated by \cite{Wilkins2021} our FRG results are more accurate as $\hat{H}^2$ is increased which corresponds to the strength of the stochastic fluctuations increasing. It is also clear that we are capturing non-trivial effects as the deviation from the bare potential prediction is significant. However this does point to a degeneracy in our results, and theoretical predictions for observations in general. \\

The values we obtained in Table~\ref{tabel:spectral_tilt} could have just as easily been obtained from a harmonic potential with a suitably modified coefficient. In this way many \emph{different} potentials can give \emph{identical} predictions for the spectral tilt. To make this point more transparent we plot in Fig.~\ref{fig:spectral_tilt_compare} the harmonic potentials $U(\sigma) \propto \sigma^2$ that would reproduce the spectral tilt predictions in Table~\ref{tabel:spectral_tilt} for the FRG methods and direct numerical simulation. We can see that at $\hat{H}^2 = 5$ that these harmonic potentials closely match the original Langevin potential which makes sense from the results in Table~\ref{tabel:spectral_tilt}. However as $\hat{H}^2$ decreases we can see that the deviations becomes more significant so that it does not resemble the original Langevin potential at all. In this way it is clear that one should be careful about making inferences about the potential from values of the spectral tilt. At high $\hat{H}^2$ one could very easily add features like bumps that would negligibly change the spectral tilt but make the potential look very different. At lower $\hat{H}^2$, added features will modify the spectral tilt significantly, but would still match the prediction from a suitably modified harmonic potential. At very low $\hat{H}^2$ however the equilibrium distribution will be heavily contained near the equilibrium point and features further away will again have little impact on the spectral tilt. \\

The values for the spectral tilt we show in Table~\ref{tabel:spectral_tilt} are too high to correspond to the curvature perturbation so the potentials we consider here could not correspond to the \emph{curvaton} scenario. While the work done in section \ref{sec:powerspec_equi} has been done assuming an equilibrium distribution, this procedure could easily be applied out of equilibrium. Section \ref{sec:two-point} yields numerically the temporal two-point function and using equations (\ref{eq:equal_time=equal_space}) and (\ref{eq:power_spec_sigma_defn}) the power spectrum can also be numerically obtained.

%\subsection{Power Spectrum from non-equilibrium evolution}
%Naturally if $\sigma$ is not in its equilibrium configuration things become more complicated. 

\section{\label{sec:FPT_spec} First-Passage Time prediction}
The FRG naturally gives us a prediction for the evolution of $\Sigma$ and the connected two point function of the field $G$ and we showed in section \ref{sec:Cosmo_Spec} how this can be turned into predictions for cosmological observables such as the power spectrum. In this section we will instead examine the FPT problem in line with our previous computations for the inflaton \cite{Rigopoulos2021}. In particular what we wish to know is the probability distribution, $\rho (\mathcal{N})$ for the number of e-folds it takes to reach a field value $\sigma_2$ given it was initially at $\sigma_1$ at some initial time $\mathcal{N}_i$ which for simplicity we identify with $0$. For a spectator field $\sigma$ that obeys the Langevin equation (\ref{eq:Langevin_sigma_dimless}) we can straightforwardly write down the Fokker-Planck (F-P) equation that the PDF $P(\sigma, \alpha)$ obeys:
\begin{eqnarray}
	\dfrac{\partial P(\sigma,\alpha)}{\partial \alpha} = \partial_{\sigma}(P(\sigma,t)\partial_{\sigma} \hat{U}) + \dfrac{\hat{H}^2}{2}\partial_{\sigma\sigma} P(\sigma,\alpha) \label{eq: F-P_spect}
\end{eqnarray}
The question now is how one can compute the solution to (\ref{eq: F-P_spect}) using FRG techniques.
\subsection{Normal Distribution}
It is true that in general the solution to (\ref{eq: F-P_spect}) is not a normal distribution, however we will assume it is as the FRG is able to accurately predict the evolution of the average position $\Sigma (\alpha)$ and the variance $G(\alpha )$ -- now denoted with a single argument for notational brevity. This gives us the following ansatz:
\begin{eqnarray}
P(\sigma, \alpha) = \dfrac{1}{\sqrt{2\pi G (\alpha)}} \exp \lsb -\dfrac{1}{2}\dfrac{\lb \sigma -\Sigma (\alpha)\rb^2}{G(\alpha)}\rsb \label{eq:pdf_gaus_spect}
\end{eqnarray}
We now wish to compute the probability $\rho (\mathcal{N})$ that $\sigma_2$ is reached between $\mathcal{N}$ and $\mathcal{N} + \mathrm{d}\mathcal{N}$ e-folds. This can be related to (\ref{eq:pdf_gaus_spect}) using \cite{VanKampen2007}:
\begin{eqnarray}
\int_{\mathcal{N}}^{\infty} \rho (\alpha) \mathrm{d}\alpha &=& \int_{\sigma_2}^{\infty}P(\sigma,\mathcal{N})\mathrm{d}\sigma \label{eq:FPT_FP_pdf_equivalence_spect_int}\\
\Rightarrow \rho (\mathcal{N}) &=& -\dfrac{\partial}{\partial \mathcal{N}} \int_{\sigma_2}^{\infty}P(\sigma,\mathcal{N})\mathrm{d}\sigma \label{eq:FPT_FP_pdf_equivalence_spect}
\end{eqnarray}
To see why this is the case consider that the LHS of (\ref{eq:FPT_FP_pdf_equivalence_spect_int}) is simply the probability that it takes longer than $\mathcal{N}$ e-folds for the field to reach $\sigma_2$. This is equal to the probability that the trajectory lasts longer than $\mathcal{N}$, $P($ trajectory $ > \mathcal{N})$. The RHS is the area under the F-P PDF above the exit point $\sigma_2$ at the time $\mathcal{N}$. Provided there is an absorbing boundary condition at $\sigma_2$ then this area is the fraction of trajectories that have not yet reached $\sigma_2$ at time $\mathcal{N}$. This means that all of these trajectories will take longer than $\mathcal{N}$ to reach $\sigma_2$ so this area does indeed equal $P($ trajectory $ > \mathcal{N})$ so the LHS = RHS. If there is not an absorbing boundary condition than the RHS will be larger than the LHS. This is because the RHS will now include contributions from trajectories that have reached $\sigma_2$ previously but are now at $\sigma > \sigma_2$ meaning we are no longer computing a true \emph{first}-passage time quantity. Therefore if one does not include an absorbing boundary condition, a computation of the RHS would \emph{overestimate} the number of trajectories that have yet to reach $\sigma_2$ and the prediction for $\rho (\mathcal{N})$ would have a fatter tail than the true value. An absorbing boundary condition is naturally imposed for the F-P equation for the inflaton -- as it corresponds to inflation ending -- but this is not so for a spectator field and the PDF (\ref{eq:pdf_gaus_spect}) is not endowed with such boundary conditions. Instead the PDF that enters (\ref{eq:FPT_FP_pdf_equivalence_spect}) will be different to the one that solves (\ref{eq: F-P_spect}). We can easily modify the ansatz (\ref{eq:pdf_gaus_spect}) to include an absorbing boundary condition by adding another Gaussian solution that cancels at $\sigma_2$:
\begin{eqnarray}
    P(\sigma, \alpha) = \dfrac{A}{\sqrt{2\pi G (\alpha)}} \lcb \exp \lsb -\dfrac{1}{2}\dfrac{\lb \sigma -\Sigma (\alpha)\rb^2}{G(\alpha)}\rsb - \exp \lsb -\dfrac{1}{2}\dfrac{\lb 2\sigma_2 -\sigma -\Sigma (\alpha)\rb^2}{G(\alpha)}\rsb\rcb \label{eq:pdf_gaus_spect_absorb}
\end{eqnarray}
Where $A$ is a normalisation factor to be determined. Substituting (\ref{eq:pdf_gaus_spect_absorb}) into (\ref{eq:FPT_FP_pdf_equivalence_spect}) yields:
\begin{eqnarray}
\rho (\mathcal{N}) = -A\dfrac{\partial}{\partial \mathcal{N}}\lsb  \text{erfc}\lb \dfrac{\sigma_2 -\Sigma (\mathcal{N})}{\sqrt{2G(\mathcal{N})}}\rb \rsb \end{eqnarray}
Which can be straightforwardly evaluated to give the main result of this paper:
\begin{eqnarray}
\rho (\mathcal{N}) = \dfrac{A}{\sqrt{2\pi G (\mathcal{N})}}\lsb \dfrac{\lb \Sigma (\mathcal{N}) - \sigma_2\rb\partial_{\mathcal{N}}G(\mathcal{N})}{2G(\mathcal{N})} - \partial_{\mathcal{N}}\Sigma (\mathcal{N})\rsb \exp \lsb -\dfrac{1}{2} \dfrac{\lb\sigma_2 -\Sigma (\mathcal{N})\rb^2}{G(\mathcal{N})}\rsb \label{eq:rhoN_spect}
\end{eqnarray}
For $\rho (\mathcal{N})$ to be a probability its integral between the initial e-fold time $\mathcal{N}_{in}$ and $\mathcal{N} \rightarrow \infty$ must be unity. This tells us that $A$ is given by:
\begin{eqnarray}
\dfrac{1}{A}=   \text{erf}\lb \dfrac{\sigma_2 -\Sigma_{in}}{\sqrt{2G_{in}}} \rb - \text{erf}\lb \dfrac{\sigma_2 -\Sigma_{eq}}{\sqrt{2G_{eq}} }\rb  \label{eq:norm_spect}
\end{eqnarray}
where subscripts $in$ and $eq$ indicate quantities evaluated at the initial e-fold time $\mathcal{N}_{in}$ and at equilibrium respectively. It is worth noting that if our initial condition corresponds to a delta function then (\ref{eq:norm_spect}) simplifies to:
\begin{eqnarray}
\dfrac{1}{A} =  1 -  \text{erf}\lb \dfrac{\sigma_2 -\Sigma_{eq}}{\sqrt{2G_{eq}} }\rb    \label{eq:norm_spect_delta}
\end{eqnarray}
and if $\sigma_2$ is the equilibrium point the norm can be further simplified to $A = 1$. Equations (\ref{eq:rhoN_spect}) \& (\ref{eq:norm_spect}) are the main results of this section.\\

We also recall from \cite{Wilkins2021} that the effective dynamical potential can be approximated by the Boltzmann potential:
\begin{eqnarray}
\tilde{U}_{Boltz}(\Sigma) = \dfrac{\hat{H}^2}{4G_{eq}}\left( \Sigma - \Sigma_{eq}\right)^2 \label{eq: BoltztildeU}
\end{eqnarray}
which suggests the effective dynamical potential is simply a harmonic potential centred at the equilibrium point with mass determined by the equilibrium variance. A potential of this form gives the following simple predictions for the average position and variance:
\begin{eqnarray}
\Sigma_{Boltz} (\alpha) &=& \Sigma_{in}\exp \lb -\dfrac{\hat{H}^2}{2G_{eq}} \alpha\rb + \Sigma_{eq}\\
G_{Boltz} (\alpha ) &=& \lsb G_{in} - G_{eq} \rsb \exp \lb -\dfrac{\hat{H}^2}{G_{eq}} \alpha\rb + G_{eq}
\end{eqnarray}
these equations can be substituted into (\ref{eq:rhoN_spect}) to give a prediction for FPT quantities and will act as a benchmark for the FRG.\\
\begin{figure}[t!]
    \centering
    \includegraphics[width = 0.45\linewidth]{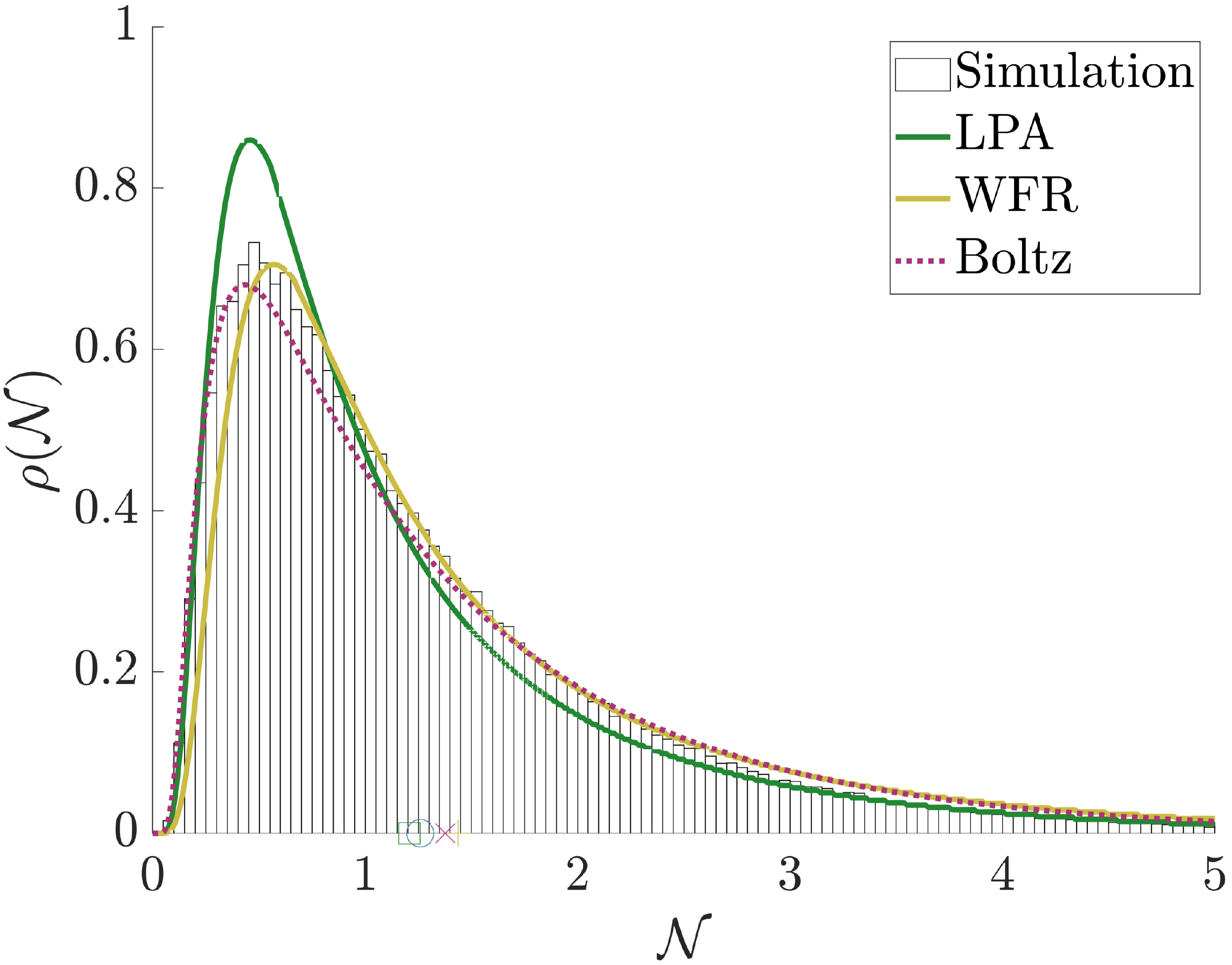}
    \includegraphics[width = 0.45\linewidth]{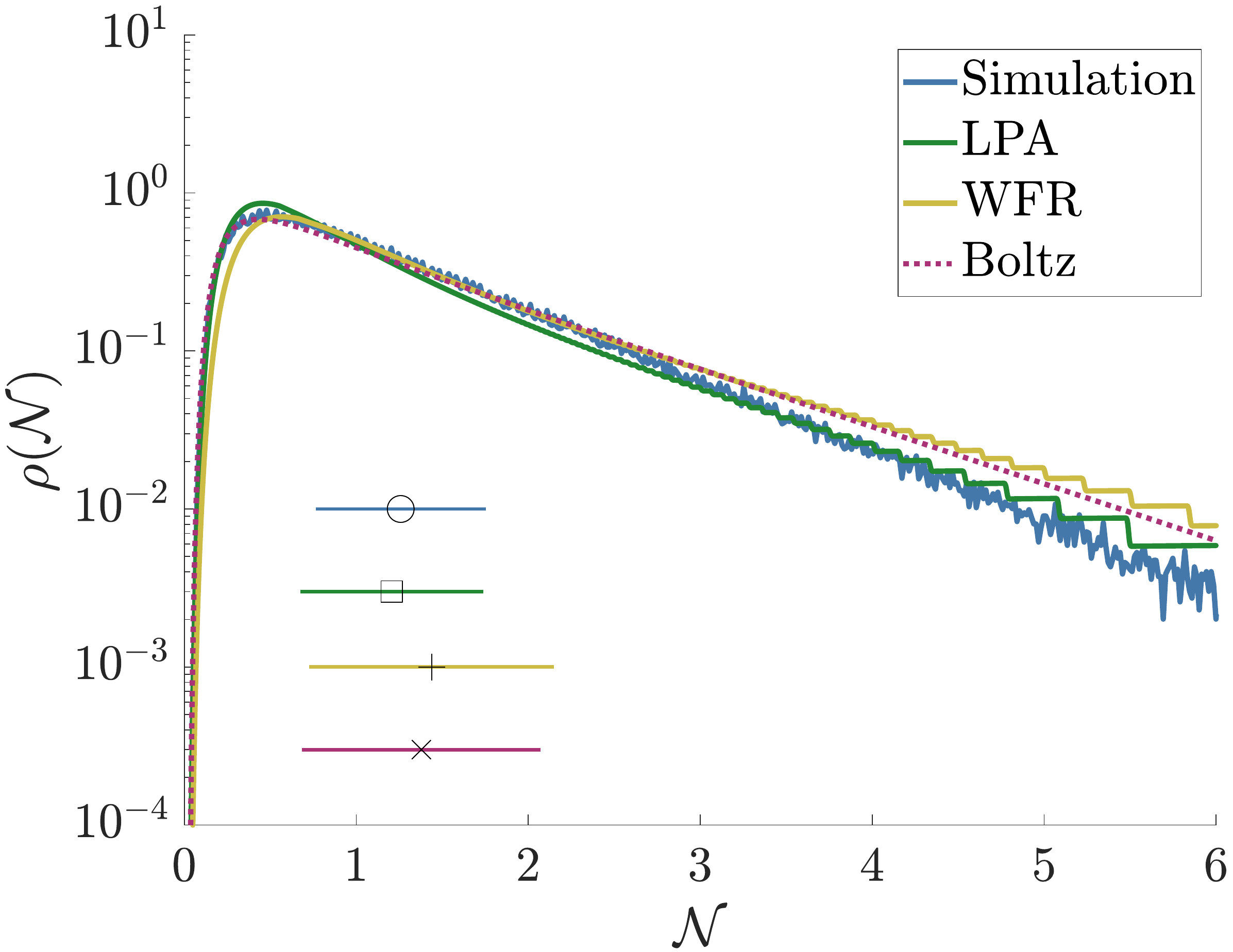}
    \includegraphics[width = 0.45\linewidth]{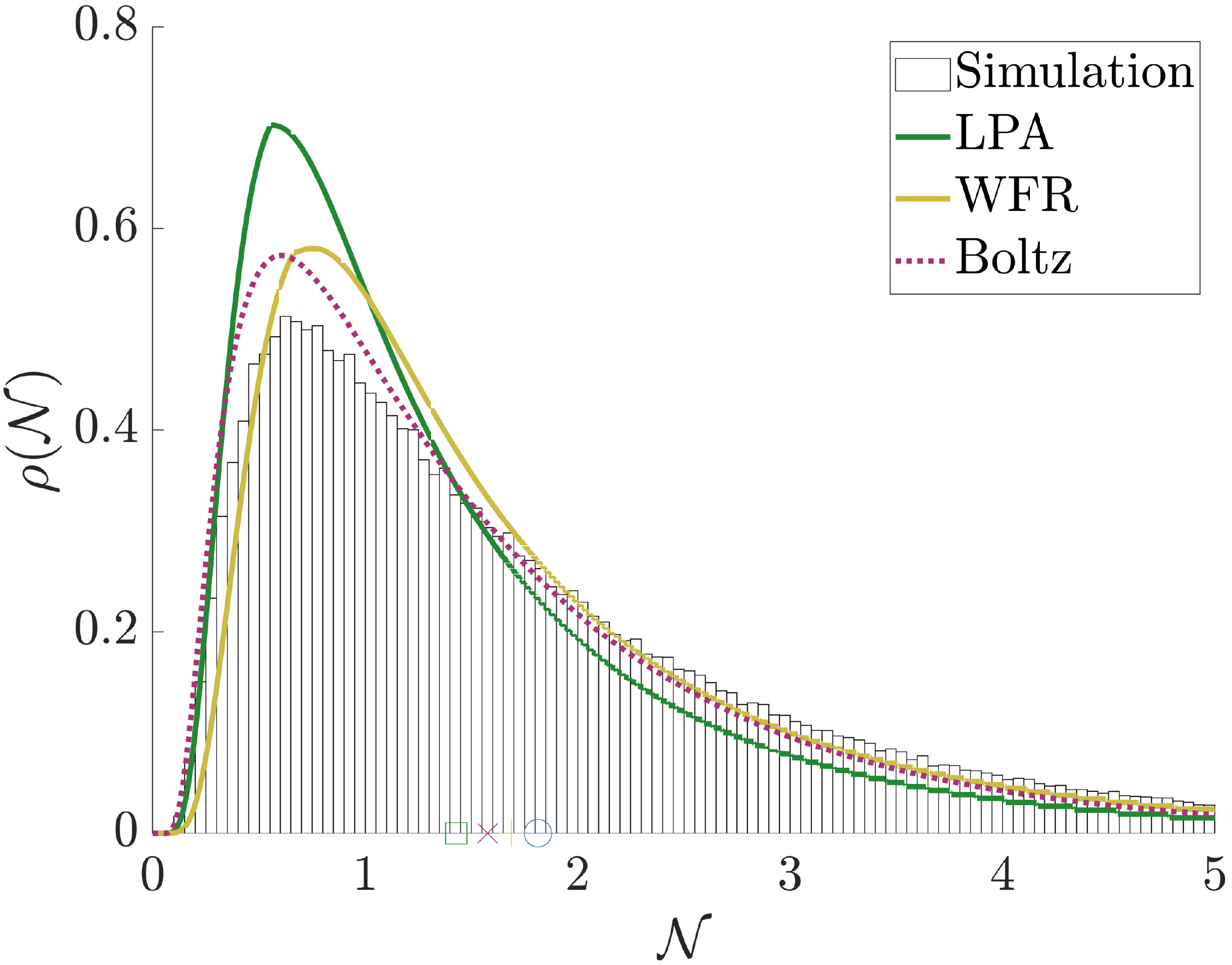}
    \includegraphics[width = 0.45\linewidth]{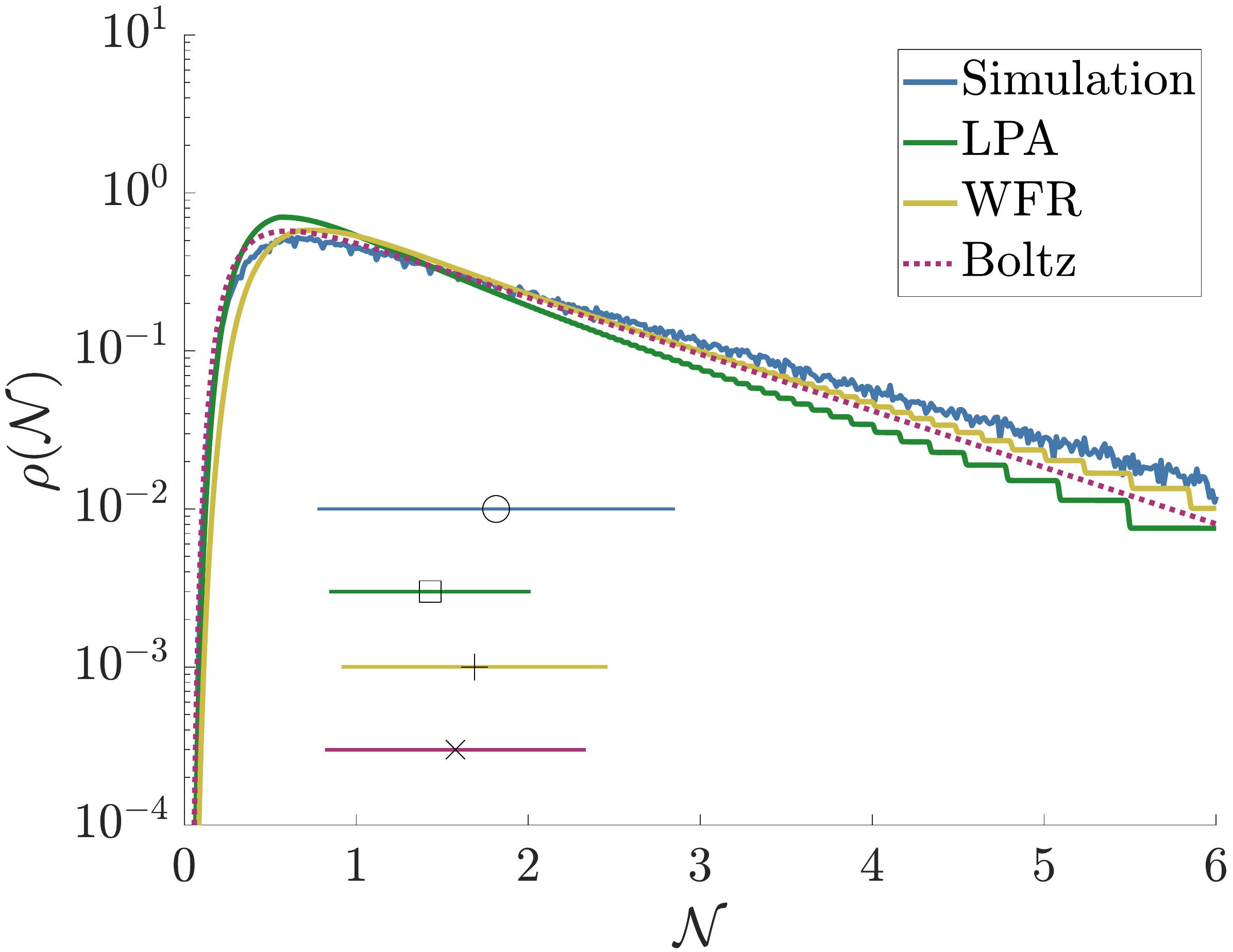}
    \caption[First-Passage Time PDF for the doublewell]{The PDF for time taken for the spectator to reach $\sigma = 0.5$ (top row) and the equilibrium $\sigma = 0$ (bottom row) in the doublewell potential for $\hat{H}^2 = 5$ using linear (left) and log (right) scales. The circle, box, plus and cross symbols in the plots represent the mean time taken $\lan \mathcal{N} \ran$ as computed by direct numerical simulation, by the LPA, by WFR and assuming a Boltzmann type potential respectively. The width of the horizontal lines in the log plots correspond to the respective variances $\delta \mathcal{N}^2 = \lan \mathcal{N}^2\ran - \lan \mathcal{N}\ran^2$ with an arbitrary vertical offset for legibility. The initial conditions were a normal distribution centred at $\sigma = 3$ with variance $ = 0.05$.}
    \label{fig:FPT5H2_DW}
\end{figure}
In Fig.~\ref{fig:FPT5H2_DW} we plot the PDF for the FPT to reach $\sigma = 0.5$ (top row) and $\sigma = 0$ (bottom row) for the doublewell potential at $\hat{H}^2 =5$ for an initially normal distribution centred at $\sigma =3$. We have compared the results from simulations with the FRG from (\ref{eq:rhoN_spect}) as well as the Boltzmann potential prediction. We can readily see -- as expected -- that WFR offers an improvement over LPA with matching the FRG. What is more surprising is how well the Boltzmann potential prediction also does even when the final position is not the equilibrium point. It is worth emphasising that the bottom row is essentially a barrier escape problem suggesting the applicability of this technique to general thermal systems.\\
\begin{figure}[t!]
    \centering
    \includegraphics[width = 0.45\linewidth]{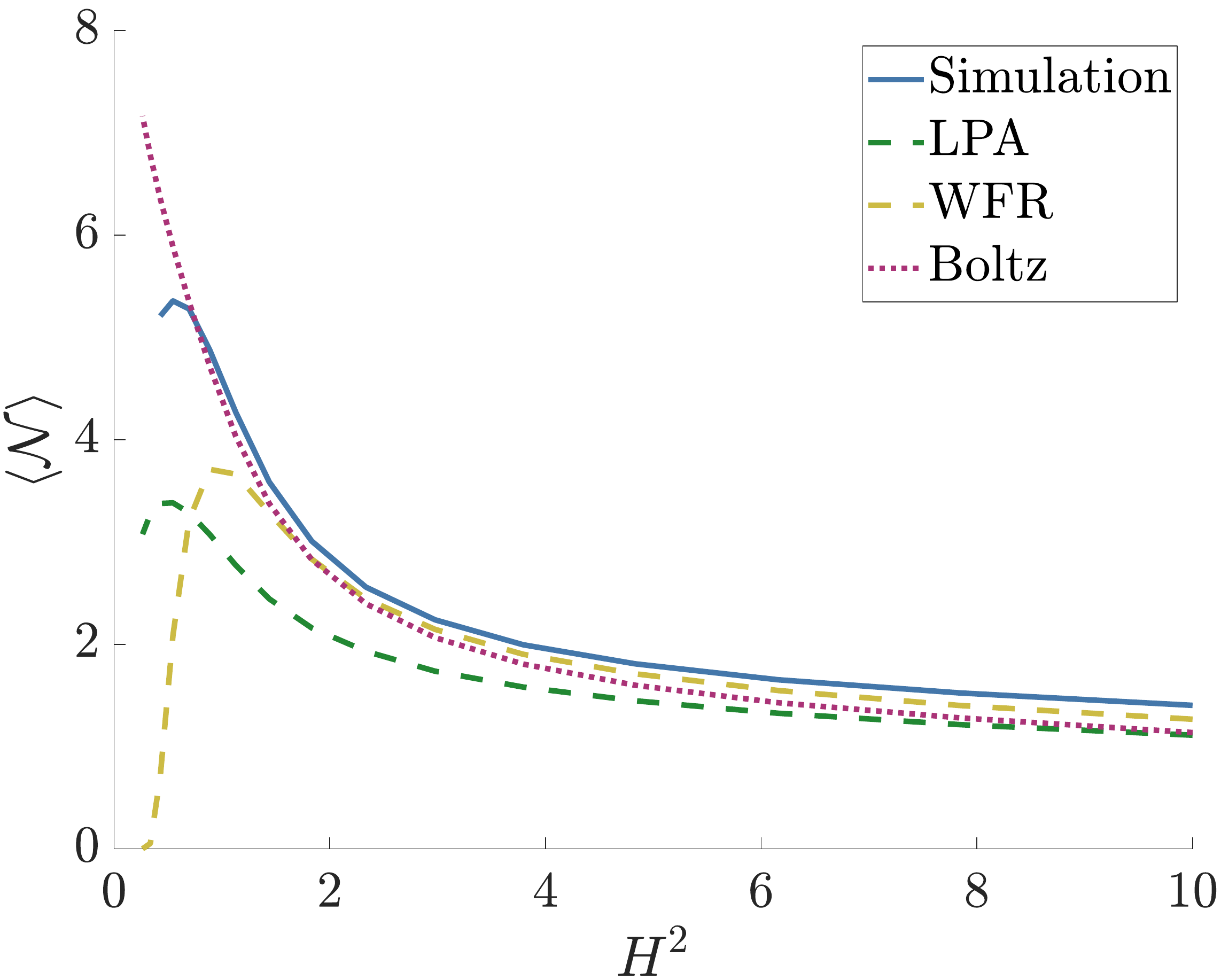}
    \includegraphics[width = 0.45\linewidth]{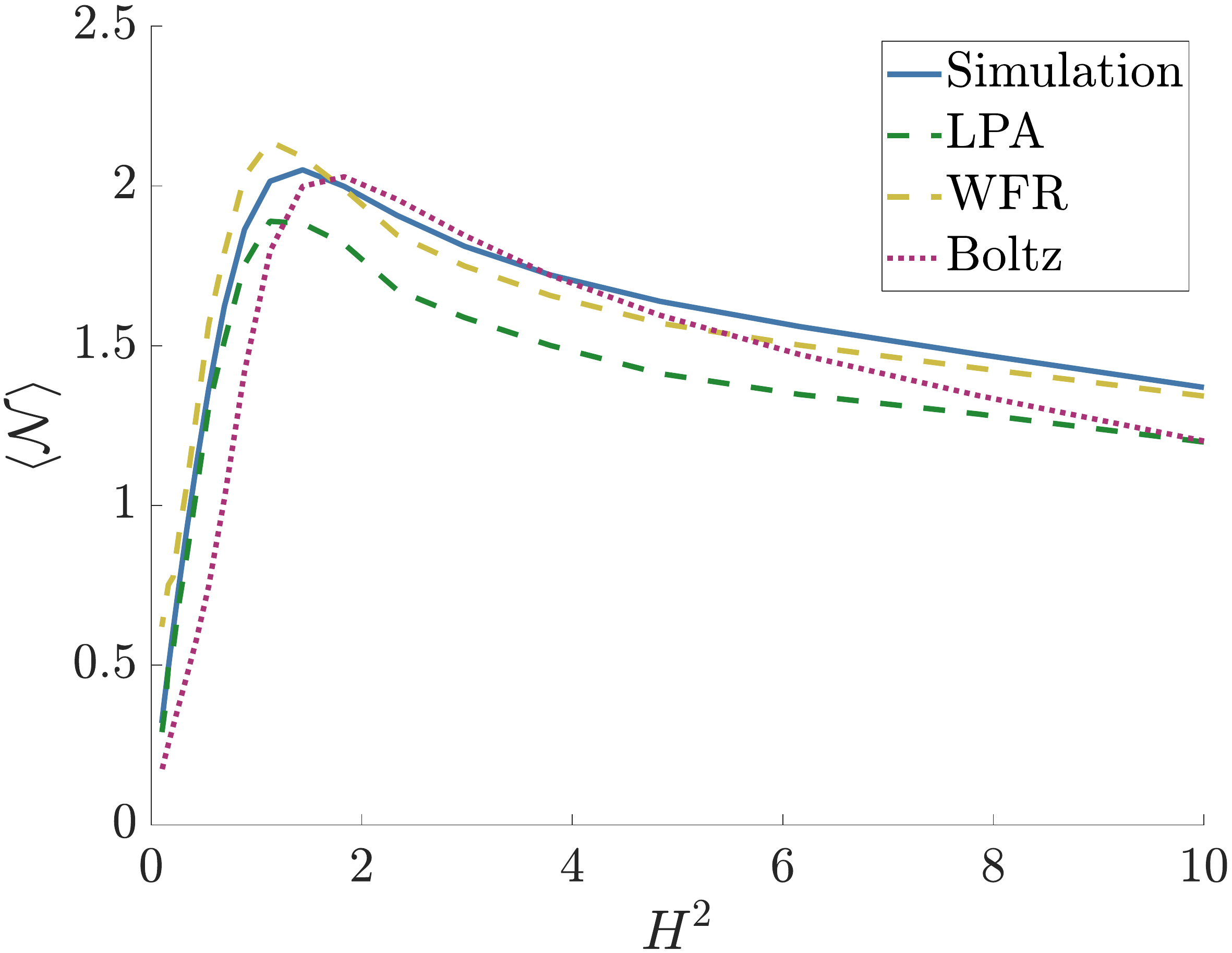}
    \caption[Average time taken to reach equilibrium for a spectator field]{Dependence of the average time taken to reach the equilibrium point, $\lan \mathcal{N} \ran$, on $\hat{H}^2$ for the doublewell potential (left) and polynomial (right) as computed by different approaches. The initial conditions were a normal distribution centred at $\sigma = 3$ with variance $ = 0.05$.}
    \label{fig:FPTALL_mean}
\end{figure}

To get a more general sense of how well the FRG does at predicting FPT quantities we plot the predictions for the average time taken, $\lan \mathcal{N} \ran $ to reach the equilibrium point for the doublewell and polynomial potentials in Fig.~\ref{fig:FPTALL_mean} over a range of $\hat{H}^2$. We can see that the FRG does a good job at correctly predicting how $\lan \mathcal{N} \ran $ changes as the value of $\hat{H}^2$ is varied with WFR in particular offers good agreement with the result from direct numerical simulation. \\

\begin{figure}[t!]
    \centering
    \includegraphics[width = 0.45\linewidth]{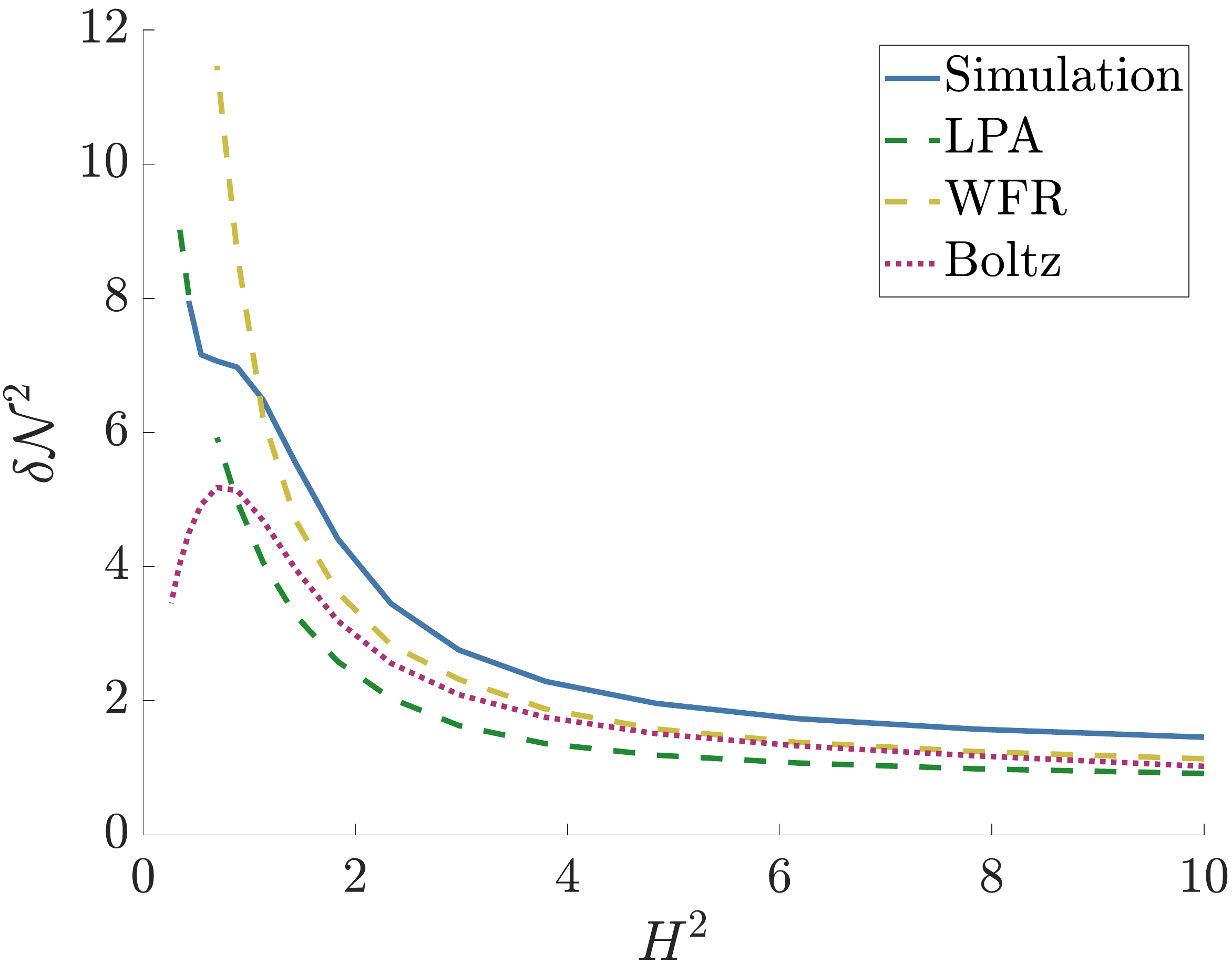}
    \includegraphics[width = 0.45\linewidth]{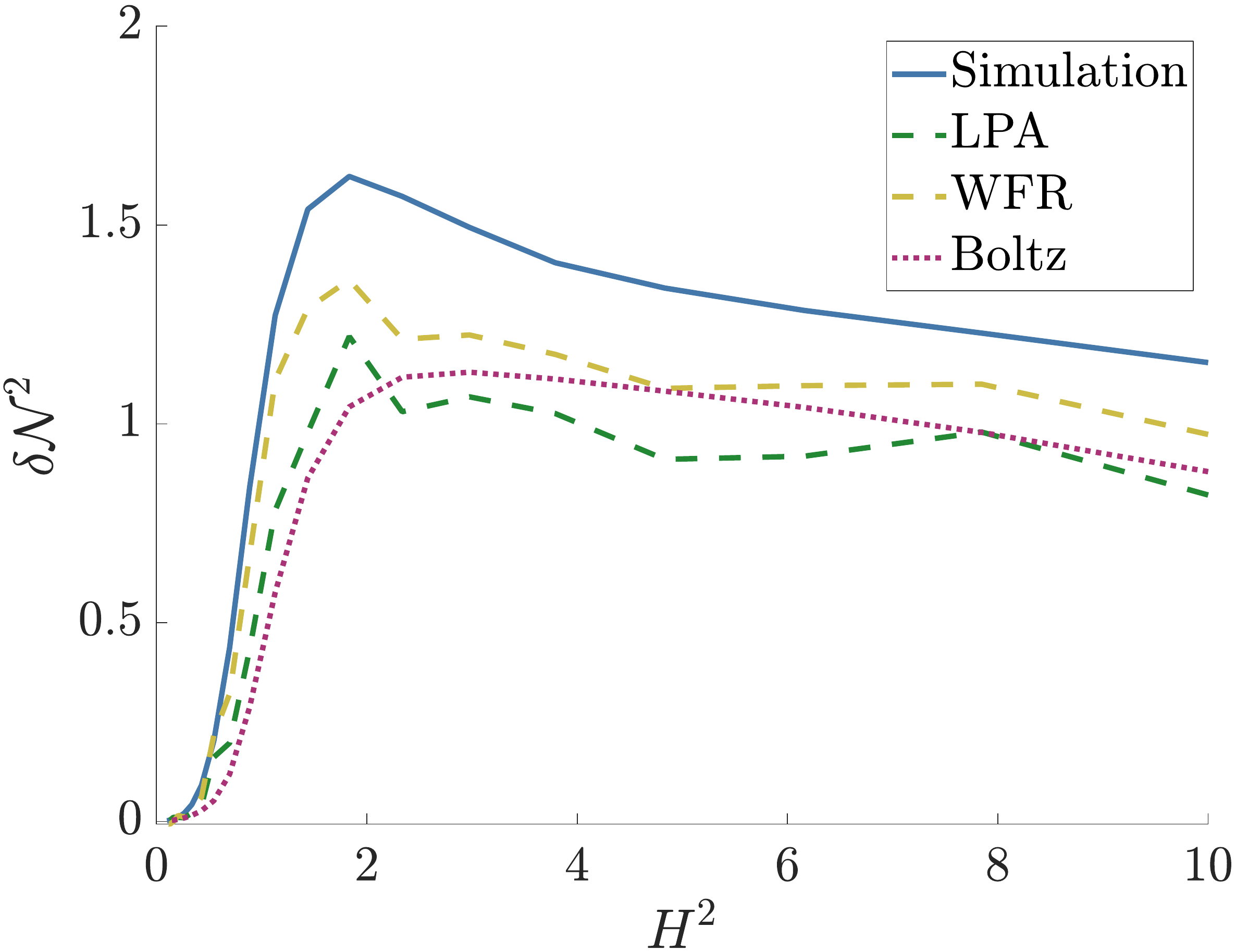}
    \caption[Variance in time taken to reach equilibrium for a spectator field]{Dependence of the variance in time taken to reach the equilibrium point, $\delta \mathcal{N}^2 = \lan \mathcal{N}^2\ran - \lan \mathcal{N} \ran^2 $, on $\hat{H}^2$ for the doublewell potential (left) and polynomial (right) as computed by different approaches. The initial conditions were a normal distribution centred at $\sigma = 3$ with variance $ = 0.05$.}
    \label{fig:FPTALL_Var}
\end{figure}
We have also plotted in Fig.~\ref{fig:FPTALL_Var} how the variance in time taken to reach equilibrium $\delta \mathcal{N}^2 = \lan \mathcal{N}^2\ran - \lan \mathcal{N} \ran^2 $ changes as the value of $\hat{H}^2$ is varied. We can see that while the FRG does not match as well as it does for $\lan \mathcal{N} \ran $  it still offers good agreement and improvement over the Boltzmann prediction. It is remarkable given the number of assumptions that had to be taken to achieve this result -- derivative expansion of the REA, simple regulator, Gaussian solution to the F-P equation -- that the FRG agrees as well as it does. \\

\section{\label{sec:RGSpec_conc}Conclusions}
In this paper we have applied the Functional Renormalisation Group (FRG) techniques developed in \cite{Wilkins2021} to the first-passage time (FPT) problem. Having outlined how the FRG equations derived in \cite{Synatschke2009,Wilkins2021} can be successfully applied to a spectator scalar field during inflation we derived the effective equations of motion (EEOM) for the third central moment in an attempt to go beyond Gaussian statistics. Unfortunately while the FRG is capable of correctly describing the qualitative nature of the third central moment it does not offer sufficient quantitative accuracy. We surmised that this is probably the limit of the accuracy of the derivative expansion of the Regulated Effective Action (REA) and that going to higher orders or focusing on a vertex expansion might offer better results. \\

\noindent We went on to discuss what cosmological observables could be predicted from an FRG approach. In the curvaton scenario the spectator field could provide the dominant contribution to the primordial curvature perturbation and therefore one may wish to compute the power spectrum and spectral tilt of $\sigma$. We reviewed how de Sitter invariance allows us to relate correlations in space -- what we observe in the CMB -- to correlations in time -- what can be computed in a stochastic approach. As the FRG predicts that (in equilibrium) the covariance follows a simple exponential in time, i.e. $\lan \sigma_0 \sigma_{\alpha}\ran \propto e^{-\lambda \alpha}$, the real space correlator follows a simple power law form. This means that the spectral tilt is simply given by $n_{\sigma} - 1 = 2\lambda$, and we showed for a $\sigma^2$ plus bumps potential that the FRG (and in particular WFR) can accurately compute the spectral tilt. We showed in Fig.~\ref{fig:spectral_tilt_compare} how this creates a degeneracy in predictions such that potentials with features, like Gaussian bumps, give the same predictions for the spectral tilt as an appropriately scaled harmonic potential. One should therefore be wary about making inferences about the potential from observational measurements like the spectral tilt. We also used the FRG to confirm the erasure of initial condition dependence of the spectator field during inflation as to be expected by the presence of the SR attracter. \\

\noindent We finished this paper by an examination of the FPT problem for a spectator field. In particular we derived an analytic formula for the PDF for time taken to traverse between two points assuming a normal distribution (\ref{eq:rhoN_spect}) -- and did the same for a skew-normal distribution (\ref{eq:rhoN_spect_skew}) in the appendix. As the FRG is able to predict the evolution of average position $\Sigma$ and its variance with time this meant the FRG could make predictions for FPT quantities. We showed that the FRG captured the shapes of the PDFs well and commented on the surprising robustness of the simple Boltzmann equilibrium prediction even far from equilibrium. We showed that even assuming a normal distribution that the FRG is able to accurately predict the average time taken to traverse between two points $\lan \mathcal{N} \ran $ and the variance in the time taken $\delta \mathcal{N}^2 = \lan \mathcal{N}^2\ran - \lan \mathcal{N} \ran^2 $ for complicated potentials where there is a barrier to be overcome. It is also capable of capturing the correct non-Gaussian tails one would expect to see in this sort of stochastic problem. This represents a first, crucial step towards using FRG techniques to compute FPT quantities for the inflaton. This also suggests that the EEOM are a useful way to predict barrier escape in thermal systems in general without having to run many costly simulations.

\appendix
\section{Skew-Normal Distribution}
In principle one can go beyond an initially Gaussian distribution and introduce skewness through the third central moment $\lan \sigma(\alpha)^3\ran_{C}$. There are many different distributions with skew but here we will assume a skew-normal distribution given by:
\begin{eqnarray}
P(\sigma, \alpha) = \dfrac{1}{\sqrt{2\pi B (\alpha)}} \lsb 1 + \text{erf}\lb \dfrac{C(\alpha)\lb \sigma - A(\alpha) \rb}{\sqrt{2B(\alpha )}}\rb \rsb \exp \lsb -\dfrac{1}{2}\dfrac{\lb \sigma -A (\alpha)\rb^2}{B(\alpha)}\rsb 
\end{eqnarray}
where the time dependent parameters $A (\alpha)$, $B(\alpha)$ and $C(\alpha)$ are related to the mean, $\Sigma (\alpha)$, variance, $G(\alpha)$, and the third central moment, $\lan \sigma (\alpha)^3\ran_{C}$, in the following way:
\begin{subequations}
\begin{align}
   D(\alpha) &= \sqrt{\dfrac{\pi}{2}}\dfrac{u^{1/3}}{\sqrt{1 + u^{2/3}}}, \quad u \equiv \dfrac{2}{4-\pi}\dfrac{\lan \sigma (\alpha)^3\ran_{C}}{G(\alpha)^{3/2}}\\
   C(\alpha) &= \dfrac{D}{\sqrt{1-D^2}}\\
   B(\alpha) &= \dfrac{G(\alpha)}{1-2D^2/\pi}\\
   A(\alpha) &=  \Sigma (\alpha) - D\sqrt{\dfrac{2G(\alpha)}{\pi}}
\end{align}\label{eq:ABCD_relations_append}
\end{subequations}
Then we can proceed as in the Gaussian case:
\begin{eqnarray}
\rho (\mathcal{N}) &=& -\dfrac{\partial}{\partial \mathcal{N}}\int_{\sigma_2}^{\infty}\dfrac{\mathrm{d}\sigma}{\sqrt{2\pi B (\mathcal{N})}} \lsb 1 + \text{erf}\lb \dfrac{C(\mathcal{N})\lb \sigma - A(\mathcal{N}) \rb}{\sqrt{2B(\mathcal{N} )}}\rb \rsb \exp \lsb -\dfrac{\lb \sigma -A (\mathcal{N})\rb^2}{2B(\mathcal{N})}\rsb  \nonumber\\
&& \\
&=& -\dfrac{\partial}{\partial \mathcal{N}}\lsb  \dfrac{1}{2} \text{erfc}\lb \dfrac{\sigma_2 - A(\mathcal{N})}{\sqrt{2B(\mathcal{N})}} \rb  + 2T \lb  \dfrac{\sigma_2 - A(\mathcal{N})}{B(\mathcal{N})}, C(\mathcal{N}) \rb \rsb \\
&=& \exp \lsb - \dfrac{\lb\sigma_2 -A (\mathcal{N})\rb^2}{2B(\mathcal{N})}\rsb \Bigg\lbrace  - \dfrac{\partial_{\mathcal{N}}C(\mathcal{N})}{2\pi\lb 1 + C(\mathcal{N})^2\rb}\exp \lsb - \dfrac{C(\mathcal{N})^2\lb \sigma_2 - A(\mathcal{N})\rb^2}{2B(\mathcal{N})}\rsb \nonumber \\
&& +\dfrac{1}{\sqrt{2\pi B (\mathcal{N})}}\lsb \dfrac{\lb A (\mathcal{N}) - \sigma_2\rb\partial_{\mathcal{N}}B(\mathcal{N})}{B(\mathcal{N})} - \partial_{\mathcal{N}}A (\mathcal{N})\rsb  \text{erfc} \lb \dfrac{\sigma_2 - A(\mathcal{N})}{\sqrt{2B(\mathcal{N})}}\rb \Bigg\rbrace \label{eq:rhoN_spect_skew}
\end{eqnarray}
where we have used Owen's T function \cite{Owen1956} defined as:
\begin{eqnarray}
T(x,a) \equiv \dfrac{1}{2\pi}\int_{0}^{a}\mathrm{d}y~\dfrac{\exp \lsb -x^2 \lb 1 + y^2\rb /2\rsb}{1+y^2}
\end{eqnarray}
so that the norm is simply given as:
\begin{eqnarray}
\lan 1 \ran &=&  \dfrac{1}{2} \text{erf}\lb \dfrac{\sigma_2 - A_{eq}}{\sqrt{2B_{eq}}} \rb - \dfrac{1}{2} \text{erf}\lb \dfrac{\sigma_2 - A_{in}}{\sqrt{2B_{in}}} \rb \nonumber \\
&&+ 2T \lb  \dfrac{\sigma_2 - A_{in}}{B_{in}}, C_{in} \rb   - 2T \lb  \dfrac{\sigma_2 - A_{eq}}{B_{eq}}, C_{eq} \rb \label{eq:norm_spect_skeq}
\end{eqnarray}
However as we previously indicated the FRG poorly predicts the third central moment -- and thus the skewness. Therefore it is perhaps not surprising that when we used (\ref{eq:rhoN_spect_skew}) to the cases analysed in the main body of the text we found that going beyond a normal distribution actually \emph{worsens} our FRG predictions for the PDF. If the third central moment can be computed more accurately -- perhaps by going beyond WFR or by examining a vertex expansion of the effective action -- then it is possible that (\ref{eq:rhoN_spect_skew}) would be more accurate than the normal distribution.

\acknowledgments
AW would like to Jack Kennedy for useful discussions about stochastic processes and Archie Cable for reading through an early draft of this work. The authors would like to thank the anonymous referee for their useful comments which have improved the quality of this work.


\bibliographystyle{ieeetr85}
\bibliography{Bibliography}
\end{document}